\documentclass[1p, final]{elsarticle}

\usepackage{subfig}
\usepackage{amsmath,bm}
\usepackage{color}
\usepackage{enumerate}
\usepackage{amssymb}
\usepackage{epstopdf}
\usepackage{epsfig}
\usepackage[table,xcdraw]{xcolor}

\usepackage{lineno,hyperref}
\modulolinenumbers[5]

\journal{Journal of Computational Physics}

\begin{document}

\begin{frontmatter}

\title{Smooth Formulation for Isothermal Compositional Simulation with Improved Nonlinear Convergence}

\author{Jiamin Jiang}
\cortext[mycorrespondingauthor]{Corresponding author}
\ead{jiamin.jiang@chevron.com}

\author{Xian-Huan Wen}

\address{Chevron Energy Technology Co.}
\address{1500 Louisiana St., Houston, TX 77002, USA}

\date{July 6, 2020}

\begin{abstract}

Compositional simulation is challenging, because of highly nonlinear couplings between multi-component flow in porous media with thermodynamic phase behavior. The coupled nonlinear system is commonly solved by the fully-implicit scheme. Various compositional formulations have been proposed. However, severe convergence issues of Newton solvers can arise under the conventional formulations. Crossing phase boundaries produces kinks in discretized equations, and subsequently causing oscillations or even divergence of Newton iterations. 

The objective of this work is to develop a smooth formulation that removes all the property switches and discontinuities associated with phase changes. We show that it can be very difficult and costly to smooth the conservation equations directly. Therefore, we first reformulate the coupled system, so that the discontinuities are transferred to the phase equilibrium model. In this way a single and concise non-smooth equation is achieved and then a smoothing approximation can be made. The new formulation with a smoothing parameter provides smooth transitions of variables across all the phase regimes. In addition, we employ a continuation method where the solution progressively evolves toward the target system. 

We evaluate the efficiency of the new smooth formulation and the continuation method using several complex problems. Compared to the standard natural formulation, the developed formulation and method exhibit superior nonlinear convergence behaviors. The continuation method leads to smooth and stable iterative performance, with a negligible impact on solution accuracy. Moreover, it works robustly for a wide range of flow conditions without parameter tuning.

\end{abstract}

\end{frontmatter}

\section{Introduction}

Gas injection processes play an important role in enhanced oil recovery (EOR). Gas injection into oil reservoir involves a number of physical mechanisms that help in mobilizing and extracting oil. Depending on pressure, temperature, and fluid compositions, immiscible or miscible displacements could occur.~The physical model required to describe the mass transfer between phases is the isothermal compositional model (Coats 1980). An Equation of State (EoS) model is usually used to determine the component partitioning across phases (Michelsen 1982).

Compositional simulation continues to be a challenging problem.~Complexities are mainly due to nonlinear couplings between multi-phase multi-component flow in porous media with thermodynamic phase behavior (Alpak 2010; Voskov 2012). Several temporal discretization schemes are available to solve the compositional model (Aziz and Settari 1979). Explicit schemes pose severe restrictions on the timestep size and are impractical for large-scale detailed reservoir models, in which the Courant-Friedrichs-Lewy (CFL) numbers can vary by several orders of magnitude throughout the domain (Jenny et al. 2009).~Therefore, the fully-implicit method (FIM) is preferred in practice, with the nonlinear system solved by a Newton method. 

Various nonlinear formulations for the compositional model have been proposed (Wong et al. 1990). Two popular formulations employ two different variable sets: (1) natural variables (Coats 1980) and (2) molar variables (Acs et al.~1985). The primary unknowns for the natural-variables set include pressure, saturations, and component molar fractions.~The conservation equations and thermodynamic constraints are assembled for cells with two phases. If one of the phases disappears during a nonlinear iteration, the corresponding saturation variable is removed, and the nonlinear system is reduced to the conservation equations. This process is referred to as `variable substitution' and is an essential ingredient of the natural-variables formulation (Wong et al.~1990; Cao 2002). Under the molar formulation, one variable-set choice is the overall composition of each component.~This approach has the advantage of avoiding the need for variable substitution, since the equations and unknowns are the same for any phase state. Numerical behaviors of nonlinear formulations were investigated and compared in some recent studies (Voskov and Tchelepi 2012; Alpak and Vink 2018).~Note that Alpak and Vink (2018) devised a flexible variable-switching formulation for general thermal-compositional flow problems.~They demonstrated that the adaptive formulation effectively improved field-scale simulations of complex processes.

For large timestep size, or poor initial guess, the standard Newton method suffers from serious convergence difficulties. Obtaining a suitable initial guess for the Newton method is referred to as globalization (Knoll and Keyes 2004). Damping, or safeguarding Newton updates is a class of globalization techniques (Deuflhard 2011). One simple heuristic strategy is to apply a local damping of variable to ensure that the value change is limited to a pre-defined range (ECLIPSE 2008). Physics-based trust-region solvers were also introduced to greatly improve the nonlinear convergence of discrete transport problems (Jenny et al. 2009; Wang and Tchelepi 2013). Voskov and Tchelepi (2011) developed a nonlinear solver specifically for molar-variables formulations. Trust regions are constructed based on the flux functions along key tie-lines in compositional space. The Newton updates are controlled from crossing inflection points and phase boundaries. Although locating trust-region boundaries is straightforward in two-phase problems with simple flow physics, it can be quite difficult for general compositional models with gravity and phase changes. Moreover, computations become expensive when the flux functions vary significantly during the iterations. 

%----------------------------------------------------------------------------------------------

Recent studies revealed that the non-differentiability (kink) resulted from switching criteria in the numerical flux may cause frequent oscillations or divergence of Newton iterations (Wang and Tchelepi 2013; Lee et al. 2015; Lee and Efendiev 2016; Hamon et al. 2016; Jiang and Younis 2017). Likewise, severe convergence problems of Newton solvers can also arise under the conventional compositional formulations. This is due to the kinks produced when crossing phase boundaries.~Phase change leads to the corresponding switches in fluid properties and discretized equations.

%----------------------------------------------------------------------------------------------

Several formulations were proposed, attempting to improve the nonlinear performance of compositional simulations (Abadpour and Panfilov 2009; Lauser et al. 2011; Voskov 2012; Gharbia et al. 2015). Unified system of equations is achieved and thus the variable-substitution process is avoided. Saturations (or phase fractions) can change continuously across phase boundaries under these formulations. However, the switches of fluid properties still occur, in one form or another. The kinks in discretized equations as the essential mechanism that causes convergence difficulties are not resolved by the previous works. Consequently, their results reported insignificant improvement in nonlinear performance compared with the standard natural formulation.

%------------------------------------------------------------------

Previous studies demonstrated that a smooth numerical flux can improve Newton behaviors (Wang and Tchelepi 2013; Lee et al. 2015; Hamon et al. 2016; Jiang and Younis 2017). The objective of this work is to develop a smooth formulation that removes all the property switches and discontinuities associated with phase changes. But as we will show, it can be very difficult and costly to make smoothing approximations directly for the conservation equations under the standard formulations. Therefore, we first reformulate the coupled system, so that the discontinuities are transferred to the phase equilibrium model. In this way a single and concise non-smooth equation is achieved and then a smoothing approximation can be made. The reformulation is based on a mixed complementarity problem (MCP) proposed for the phase equilibrium in the field of chemical process simulation (Bullard and Biegler 1993; Gopal and Biegler 1997; Sahlodin et al. 2016). The MCP model contains complementarity conditions that represent phase changes. The subsequently developed formulation with a smoothing parameter can lead to smooth transitions of variables across all the phase regimes. 

For a smoothing parameter with sufficient values, the smooth coupled system exhibits superior global convergence behavior. However, solution accuracy may be degraded, with a fixed smoothing parameter. Its value needs to be adaptively determined to achieve an optimal balance between accuracy and nonlinear performance.~For more robust and general applications, we employ a continuation method with the smooth system as a homotopy mapping. The smoothing parameter is progressively reduced after each Newton iteration, and the solution evolves toward the target system. In this work, the continuation method acts as a globalization stage to obtain better initial guesses for the Newton process (Jiang and Tchelepi 2018). As will be shown, the new smooth formulation is very effective for globalizing the compositional flow problem. It is worth noting that the smooth formulation and the continuation method can be used for different applications depending on specific accuracy and implementation considerations. 

%-----------------------------------------------------------------------

We evaluate the efficiency of the new approaches using several complex examples. We focus on the nonlinear behavior of the coupled conservation and phase equilibrium system. The smooth formulation produces valid and unique solutions in all the three phase regimes. For most of the cases, the standard natural formulation suffers from multiple timestep cuts and subsequent wasted Newton iterations. In contrast, the smooth formulation and the smoothing based continuation method (SBC) exhibit superior global convergence, requiring no timestep cut. Moreover, the SBC method can largely resolve the convergence issues due to phase changes, with a negligible impact on solution accuracy. We find that applying SBC for a few iterations in the globalization stage is sufficient. In addition, the developed method works robustly for a wide range of flow conditions without parameter tuning.

\section{Isothermal compositional model}

% We apply an EoS-based compositional formulation for isothermal processes 

We consider compressible gas-oil flow in porous media without capillarity. We ignore water that does not exchange mass with the hydrocarbon phases. 

The conservation equations for the isothermal compositional problem containing $n_c$ components are written as, 
\begin{equation} 
\label{eq:mass_con_comp}
\frac{\partial}{\partial t } \left [ \phi \left ( x_c \rho_{o} s_{o} + y_c \rho_{g} s_{g} \right ) \right ] + \nabla \cdot \left ( x_c \rho_{o} \textbf{u}_{o} + y_c \rho_{g} \textbf{u}_{g} \right ) - q_{c} = 0,
\end{equation}
where $c \in \left \{ 1,...,n_c \right \}$. $x_{c}$ and $y_{c}$ are molar fractions of component $c$ in the oil and gas phases, respectively. $\phi$ is rock porosity and $t$ is time. $\rho_l$ is phase molar density. $s_{l}$ is phase saturation. $q_{c}$ is well flow rate. 

Phase velocity $\textbf{u}_l$ is expressed as a function of phase potential gradient $\nabla \Phi_l $ using the extended Darcy's law,
\begin{equation} 
\label{eq:phase_vel}
\textbf{u}_l = - k \lambda_l \nabla \Phi_l = -k\lambda_l\left ( \nabla p - \rho_l g \nabla h \right ).
\end{equation}
where $k$ is rock permeability. $p$ is pressure. Capillarity is assumed to be negligible. $g$ is gravitational acceleration and $h$ is height. Phase mobility is given as $\lambda_{l} = k_{rl}/\mu_l$. $k_{rl}$ and $\mu_l$ are relative permeability and viscosity, respectively.

Phase velocity can also be expressed under the fractional-flow formulation, 
\begin{equation} 
\textbf{u}_l = \frac{\lambda_{l}}{\lambda_T} \mathbf{u}_T + k g \nabla h \sum_{m} \frac{\lambda_{m}\lambda_{l}}{\lambda_T}  \left ( \rho_{l}-\rho_{m} \right),
\end{equation}
which involves the total velocity, 
\begin{equation} 
\label{eq:tol_vel}
\textbf{u}_T = \sum_l \textbf{u}_l = -k \lambda_T \nabla p + k \sum_l \lambda_l \rho_l g \nabla h .
\end{equation}
where the total mobility $\lambda_{T} = \sum_{l} \lambda_{l}$. In this work, Eq.~(\ref{eq:phase_vel}) is used for fully-implicit compositional simulations.

To close the nonlinear system, additional equations are needed.~These include the thermodynamic equilibrium constraints,
\begin{equation} 
\label{eq:vle_fu}
f_{c,o}(p,\textbf{x}) - f_{c,g}(p,\textbf{y}) = 0 ,
\end{equation}
where $p$, $T$, and $z_c$ denote pressure, temperature, and overall molar fraction, respectively. $f_{c,l}$ is the fugacity of component $c$ in phase $l$. 

We now write the phase constraints, 
\begin{equation} 
\label{eq:phase_const}
\sum_{c=1}^{n_c} x_{c} - 1 = 0, \qquad \sum_{c=1}^{n_c} y_{c} - 1 = 0,
\end{equation}
and the saturation constraint as,
\begin{equation} 
\label{eq:satu_const}
s_o + s_g - 1 = 0.
\end{equation}

The above system of equations provides a complete mathematical statement for two-phase multi-component flow. The local equilibrium constraints are enforced only when both phases are present.

\section{Natural-variables formulation}

% The first type of variable-set choice is the so-called ...

% There is consensus in the community that the mass conservation equations should be used as the primary equations. However, there are several choices for the primary unknowns.

An important aspect of any compositional formulation is the choice of dividing the equations and unknowns into primary and secondary sets. Two widely used formulations are (1) natural variables (Coats 1980) and (2) overall-composition variables (Collins et al. 1992; Voskov and Tchelepi 2012).

The primary unknowns include pressure, saturations, and molar fractions: 

(1) $p$ $-$ pressure [1],

(2) $s_l$ $-$ phase saturations [2], 

(3) $x_{c}$, $y_{c}$ $-$ phase compositions of each component [2$n_c$]. 
\\
The size of each variable is given in square bracket.

The various coefficients can be obtained as functions of the base variables. For a two-phase cell, the molar phase fraction is related to saturation as follows, 
\begin{equation} 
\nu_l = \frac{\rho_l s_l}{\sum_{m} \rho_m s_m}
\end{equation}
and overall molar fraction of component $c$ is written as, 
\begin{equation} 
z_c = x_{c} \nu_o + y_{c} \nu_g 
\end{equation}

Note that for single-phase ($l$) mixture, $\nu_l = s_l = 1$, and $x_{c,l} \equiv z_c$.

\subsection{Variable substitution}

%  compute the equilibrium phase behavior. 

An essential ingredient of the natural-variables formulation is the `variable substitution' process (Wong et al. 1990; Cao 2002; Voskov and Tchelepi 2012). A common strategy for variable-switching between Newton iterations during a time step is:

1. For any cell whose status in the previous iteration is single-phase, run the phase stability test (Michelsen 1982a) to check if the mixture becomes two-phase. For the mixture that splits into two phases, perform the flash to compute the phase compositions (Michelsen 1982b). 

2. If a cell is already in the two-phase state, the thermodynamic constraints are included in the nonlinear system as part of the global Jacobian.

3. If a phase saturation, or phase fraction, becomes negative between two successive iterations, the phase disappears, and appropriate variable-switching is performed. 

The system of conservation equations is solved for single-phase regimes, and the combination of conservation equations and thermodynamic constraints is solved for the two-phase regime.

\subsection{Phase behavior}

Phase behavior computation is usually a stand-alone procedure for detecting phase changes.~For a mixture of $n_c$ components and two phases, the mathematical model describing the thermodynamic equilibrium is (Voskov and Tchelepi 2012),

\begin{equation} 
\label{eq:vle_fu}
f_{c,o}(p,\textbf{x}) - f_{c,g}(p,\textbf{y}) = 0 ,
\end{equation}

\begin{equation} 
\label{eq:mass_zc}
z_c - \nu_o x_{c} - \left( 1 - \nu_o \right) y_{c} = 0 ,
\end{equation}

\begin{equation} 
\label{eq:x_y_equ}
\sum_{c=1}^{n_c} \left ( x_{c} - y_{c} \right ) = 0.
\end{equation}
where $\nu_l$ is molar fraction of phase $l$. We assume that $p$, $T$, and $z_c$ are known.~The objective is to find all the $x_{c}$, $y_{c}$ and $\nu_l$. 

Phase behavior is often described using an Equation of State (EoS) model.~EoS-based computations are expensive and may consume a large portion of total simulation time.~But their cost can be largely reduced through some advanced approaches (Voskov and Tchelepi 2009; Zaydullin et al. 2012; Yan et al. 2013).

In this work, we mainly focus on the $K$-value based method to perform phase behavior computations.~Our motivation is to isolate the nonlinear difficulties (discontinuities) specifically caused by phase boundaries.~We intend to pinpoint and analyze the associated mechanisms, without the complication from the nonlinearities of EoS models. The $K$-value method assumes that components partition across phases with fixed ratios ($K$-values). Then the fugacity constraint (\ref{eq:vle_fu}) can be rewritten as, 
\begin{equation} 
\label{eq:vle_constK}
f_{c,o} - f_{c,g} = 0 \ \ \Rightarrow \ \ y_c - K_c x_c = 0 
\end{equation}
where $K_c$ is the equilibrium ratio, which depends on pressure and temperature.

\subsection{Discontinuous issues crossing phase boundaries}

Recent studies have revealed that the non-differentiability (kink) in the numerical flux due to switching criteria can be a major cause of nonlinear convergence difficulties (Wang and Tchelepi 2013; Lee et al. 2015; Hamon et al. 2016; Jiang and Younis 2017). The curvature of residual function changes abruptly at the kink, leading to oscillations between Newton iterations (flip-flopping) or convergence failure. 

Likewise, crossing phase boundaries produces kinks in compositional models. Frequent phase changes and oscillations around phase boundaries can cause severe convergence problems (Abadpour and Panfilov 2009; Lauser et al. 2011; Voskov and Tchelepi 2011; Voskov 2012).

In a single-phase regime, the component compositions are no longer controlled by the thermodynamic equilibrium. Then Eqs. (\ref{eq:x_y_equ}) and (\ref{eq:vle_constK}) cannot be satisfied at the same time. The compositions of the existing phase can only be obtained from Eq. (\ref{eq:mass_zc}), where $\nu_l$ is set to 1 (or 0). For the missing phase, the compositions become undefined.

Phase change leads to the corresponding switches in fluid properties and discretized equations. To demonstrate this, we present the accumulation term in (\ref{eq:mass_con_comp}) for the different phase states as, 
\begin{equation} 
\label{eq:accu_disc}
\begin{cases}
x_c \rho_o (\textbf{x}) s_{o} + y_c \rho_g (\textbf{y}) s_{g} \ , \ \ \textrm{two-phase}, \\ 
z_c \rho_l (\textbf{z}) \ , \ \ \textrm{one-phase}. 
\end{cases}
\end{equation}
where $\rho_o (\textbf{x})$ indicates the density computed at a composition $\textbf{x}$, and $\rho_l (\textbf{z})$ is the density computed in the single-phase regime at a composition $\textbf{z}$.

In the absence of gravity, the overall fractional flow function of component $c$ derived from (\ref{eq:mass_con_comp}) becomes, 
\begin{equation} 
\label{eq:uc_comple}
u_c = \begin{cases}
x_c \rho_{o} \frac{\lambda_{o}}{\lambda_T} + y_c \rho_{g} \frac{\lambda_{g}}{\lambda_T} \ , \ \ \textrm{two-phase}, \\ 
z_c \rho_{l} \ , \ \ \textrm{one-phase}. 
\end{cases}
\end{equation}

By assigning unit values to $\rho_{l}$ and $\mu_l$, Eq.~(\ref{eq:uc_comple}) for component 1 can be further simplified as, 
\begin{equation} 
u_1 = \begin{cases}
x_1 \frac{k_{ro}}{k_{ro} + k_{rg}} + y_1 \frac{k_{rg}}{k_{ro} + k_{rg}} \ , \ \ \textrm{two-phase}, \\ 
z_1 \ , \ \ \textrm{one-phase}. 
\end{cases}
\end{equation}

\begin{figure}[!htb]
\centering
\includegraphics[scale=0.5]{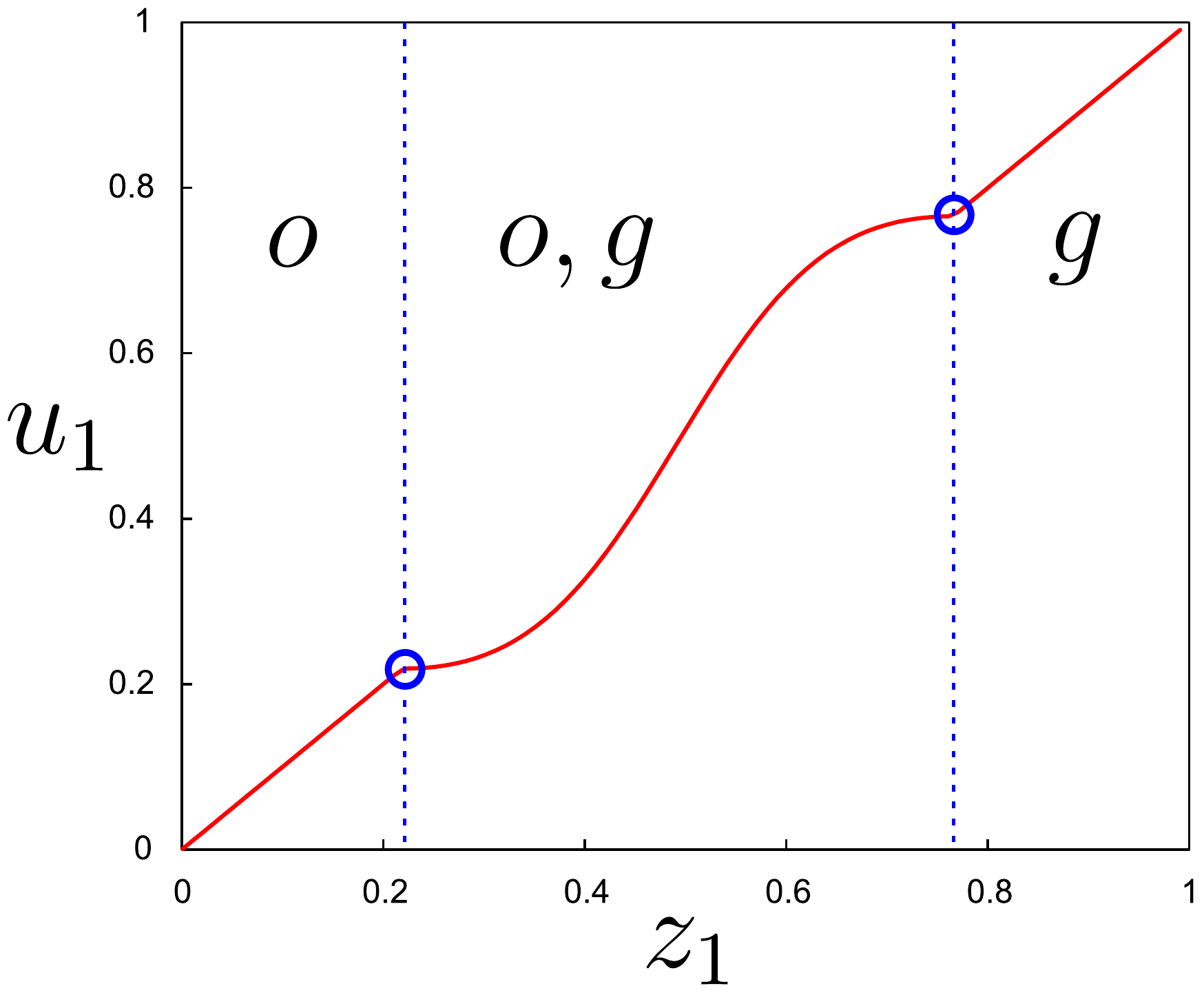}
\caption{Overall fractional flow of component $C_1$}
\label{fig:stan_F1}
\end{figure}

We consider a two-component system $\left \{ K_1 = 3.5 , K_2 = 0.3 \right \}$ and quadratic relative permeabilities with unit end points.~\textbf{Fig.~\ref{fig:stan_F1}} shows the plot of $u_1$ as a function of $z_1$. Clearly, we observe two non-differentiable points associated with the phase changes. Within the two-phase regime, the flow curve has the typical S-shape.~The flow curve becomes a straight line for the single-phase states. The discontinuous derivatives in the flow function can largely degrade the convergence performance of Newton solvers.

%===========================================================================

Several nonlinear formulations were previously proposed for compositional simulations (Abadpour and Panfilov 2009; Lauser et al. 2011; Voskov 2012; Gharbia et al. 2015). Through reformulation or certain consistence conditions, unified system of equations is achieved so that the variable-substitution process is avoided. Although saturations (or phase fractions) can change continuously across phase boundaries under these formulations, the switches of fluid properties still occur, in one form or another. The kinks in discretized equations as the essential mechanism that causes nonlinear convergence difficulties are not resolved by the previous works. Consequently, the results reported insignificant improvement in nonlinear performance compared with the conventional natural formulation.

%===========================================================================

\section{Smooth formulation}

From the previous studies (Wang and Tchelepi 2013; Lee et al. 2015; Lee and Efendiev 2016; Hamon et al. 2016; Jiang and Younis 2017), it is expected that a smooth system can provide much improved Newton behaviors.~According to Eqs.~(\ref{eq:accu_disc}) and (\ref{eq:uc_comple}) the nonlinearities (discontinuities) associated with phase changes are due to the switches of properties from the accumulation and flux terms. We can see that it is hard to develop smoothing approximations directly for those terms in the conservation equations. 

Therefore, we seek to first reformulate the coupled system, transferring the discontinuities to the phase equilibrium model. In this way a single and concise non-smooth equation is achieved and then a smoothing approximation can be derived.

\subsection{Non-smooth reformulation}

In the field of chemical process simulation, a mixed complementarity problem (MCP) was proposed for the phase equilibrium system (Bullard and Biegler 1993; Gopal and Biegler 1997; Sahlodin et al. 2016). 

The MCP model containing complementarity conditions that represent phase changes can be equivalently written in a non-smooth form as, 

\begin{equation} 
\label{eq:n_s_R_S}
R_{c} = y_c - \beta K_c x_c \ ,
\end{equation}

\begin{equation} 
R_{c + n_c} = z_c - \nu_o x_{c} - \left( 1 - \nu_o \right) y_{c} \ ,
\end{equation}

\begin{equation} 
R_{2n_c+1} = \sum_{c=1}^{n_c} \left ( x_{c} - y_{c} \right ) \ ,
\end{equation}

\begin{equation} 
\label{eq:n_s_R_E}
R_{2n_c+2} = \textrm{mid} \Big \{ \left( 1 - \nu_o \right), \left ( \beta - 1 \right ), \left ( - \nu_o \right ) \Big \} \ .
\end{equation}
where $c \in \left \{ 1,...,n_c \right \}$. The system $\textbf{R} = 0$ is simultaneously solved for the given $z_c$ and $K_c$ using a Newton method. A flash procedure that combines the Successive Substitution Iteration (SSI) and Newton method can be employed to solve the above system (Michelsen 1982b). The mid function picks the median of its three arguments, 
\begin{equation} 
\textrm{mid} \left \{ a,b,c \right \}
= \left\{\begin{matrix}
a & \textrm{if} \ c \leq a \leq b , \\
b & \textrm{if} \ c \leq b \leq a , \\
c & \textrm{if} \ b \leq c \leq a.
\end{matrix}\right.
\end{equation}
for any $\left ( a,b,c \right ) \in R^3$. 

A non-physical variable $\beta$ is introduced to relax the equilibrium equation. Then the conditions for the three phase states are given as,

\begin{equation} 
\begin{cases}
\ \beta \leq 1 \ , & \ \nu_o = 0  \\ 
\ \beta \geq 1 \ , & \ \nu_o = 1  \\ 
\ \beta = 1    \ , & \ 0 < \nu_o < 1 
\end{cases}
\end{equation}

%-------------------------------------------------------------------------------

The solutions obtained from the non-smooth system (\ref{eq:n_s_R_S} $-$ \ref{eq:n_s_R_E}) are valid and unique in all the three phase regimes. We can see that the phase fractions (thus saturations) are bounded between 0 and 1. The variables for the non-existent phase can be viewed as pseudo molar fractions that also sum to one. For the two-phase state, the non-smooth system provides the same results as the conventional phase equilibrium model.

%-------------------------------------------------------------------------------

The proposed non-smooth formulation can be readily applied to general EoS-based systems, by replacing Eq.~(\ref{eq:n_s_R_S}) with the fugacity constraint (\ref{eq:vle_fu}). $f_{c,l}$ will be governed by a nonlinear EoS model in such cases.

\subsection{Smoothing approximation}

% Chen-Harker-Kanzow-Smale smoothing functions 

To present the smooth formulation for the MCP model, we first write the mid function as a mixed complementarity function, 
\begin{equation} 
\label{eq:mcp_func}
\Theta \left ( a,b,c \right ) = \left ( a + c \right ) - \sqrt{\left ( a-b \right )^{2}} + \sqrt{\left ( b-c \right )^{2}}
\end{equation}

Then we construct a smoothing function for Eq.~(\ref{eq:mcp_func}) as follows (Chen and Harker 1993; Kanzow 1996; Li and Fukushima 2000), 
\begin{equation} 
\label{eq:mcp_smooth}
\Theta_{\epsilon} \left (\epsilon, a,b,c \right ) = \left ( a + c \right ) - \sqrt{\left ( a-b \right )^{2} + \epsilon} + \sqrt{\left ( b-c \right )^{2} + \epsilon}
\end{equation}
where $\epsilon \geq 0$ is a smoothing parameter, and $\Theta_{\epsilon} \left (\epsilon, a,b,c \right )$ is continuously differentiable with $\epsilon > 0$. We can easily see that,
\begin{equation} 
\Theta_{\epsilon} \left (0, a,b,c \right ) = \Theta \left (a,b,c \right ) = 2 \ \textrm{mid} \left \{ a,b,c \right \}
\end{equation}
whenever $a > c$.

%----------------------------------------------------------------

There are several other ways to make smoothing approximations for the mid function. We choose Eq. (\ref{eq:mcp_smooth}) because the function is computationally simple and is well-behaved in terms of nonlinearity. The phase equilibrium model with (\ref{eq:mcp_smooth}) provides smooth transitions of the variables across all the phase regimes.

%==========================================================================================

\begin{figure}[!htb]
\centering
\subfloat[Composition $y_1$ ]{
\includegraphics[scale=0.5]{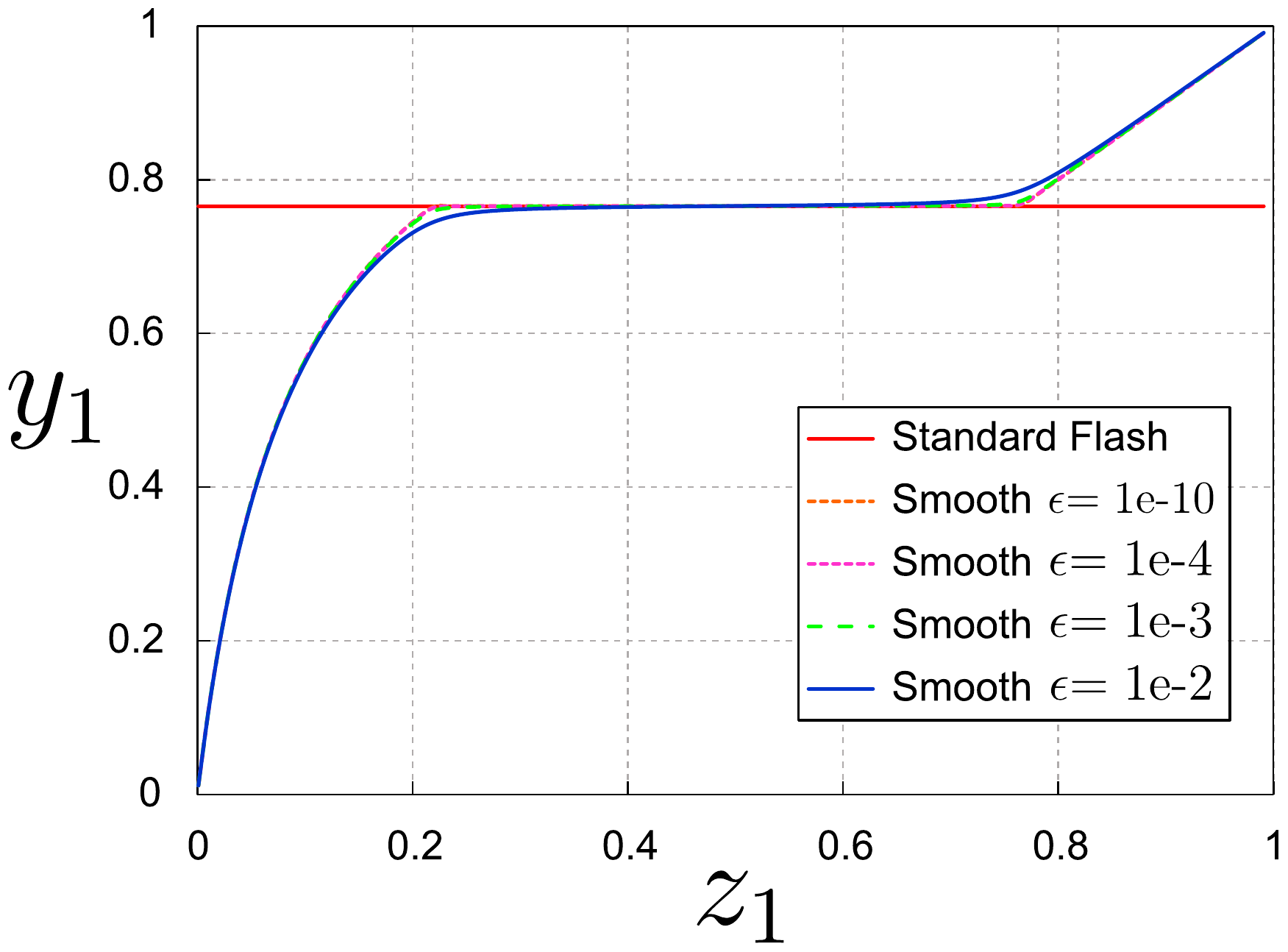}}
\\
\subfloat[Composition $x_1$ ]{
\includegraphics[scale=0.5]{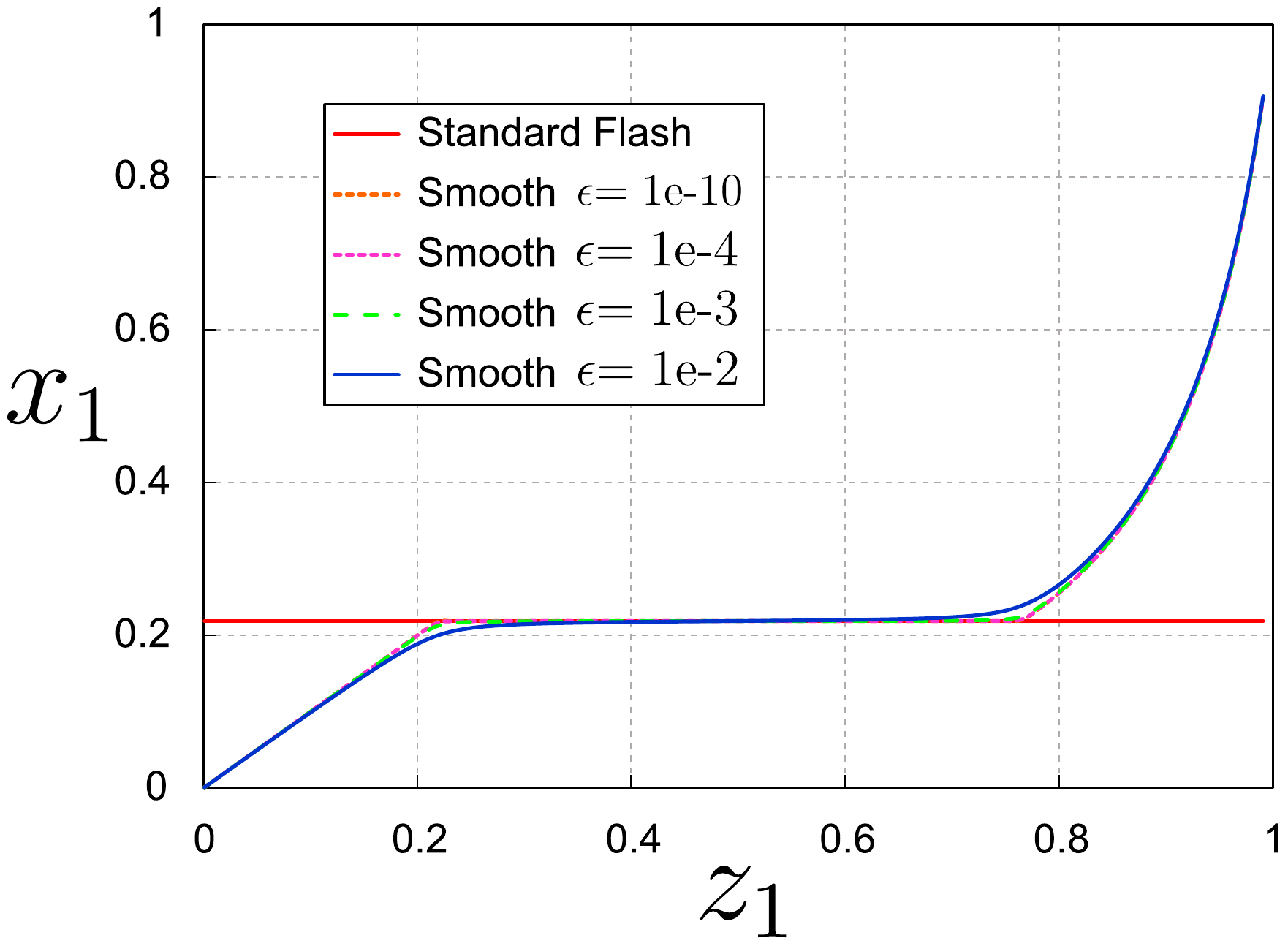}}
\caption{Compositions of the two-component fluid under the different flash formulations.}
\label{fig:2comp_g_o_comp}
\end{figure}

\begin{figure}[!htb]
\centering
\subfloat[Gas phase fraction]{
\includegraphics[scale=0.5]{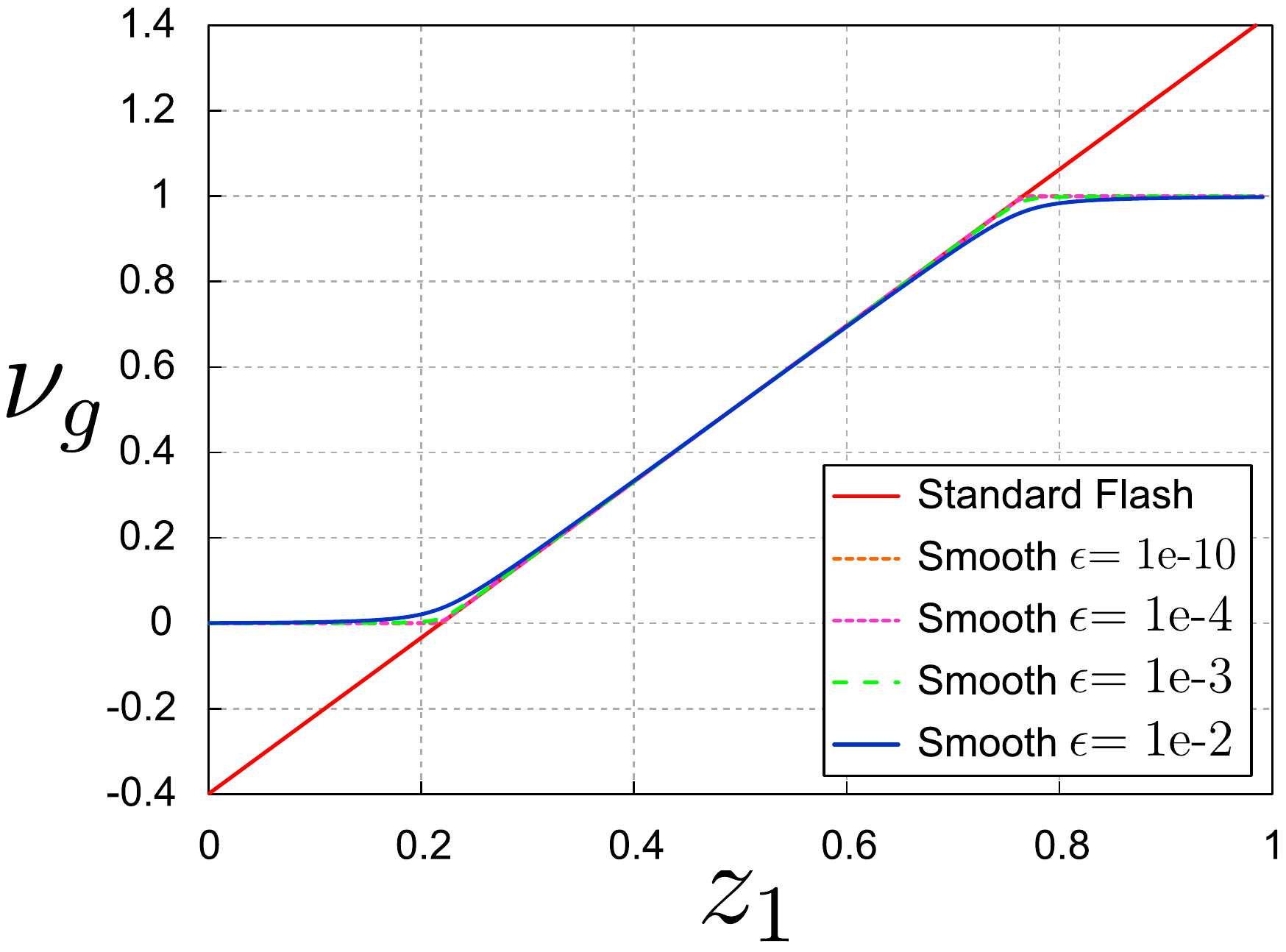}}
\\
\subfloat[Overall fractional flow]{
\includegraphics[scale=0.49]{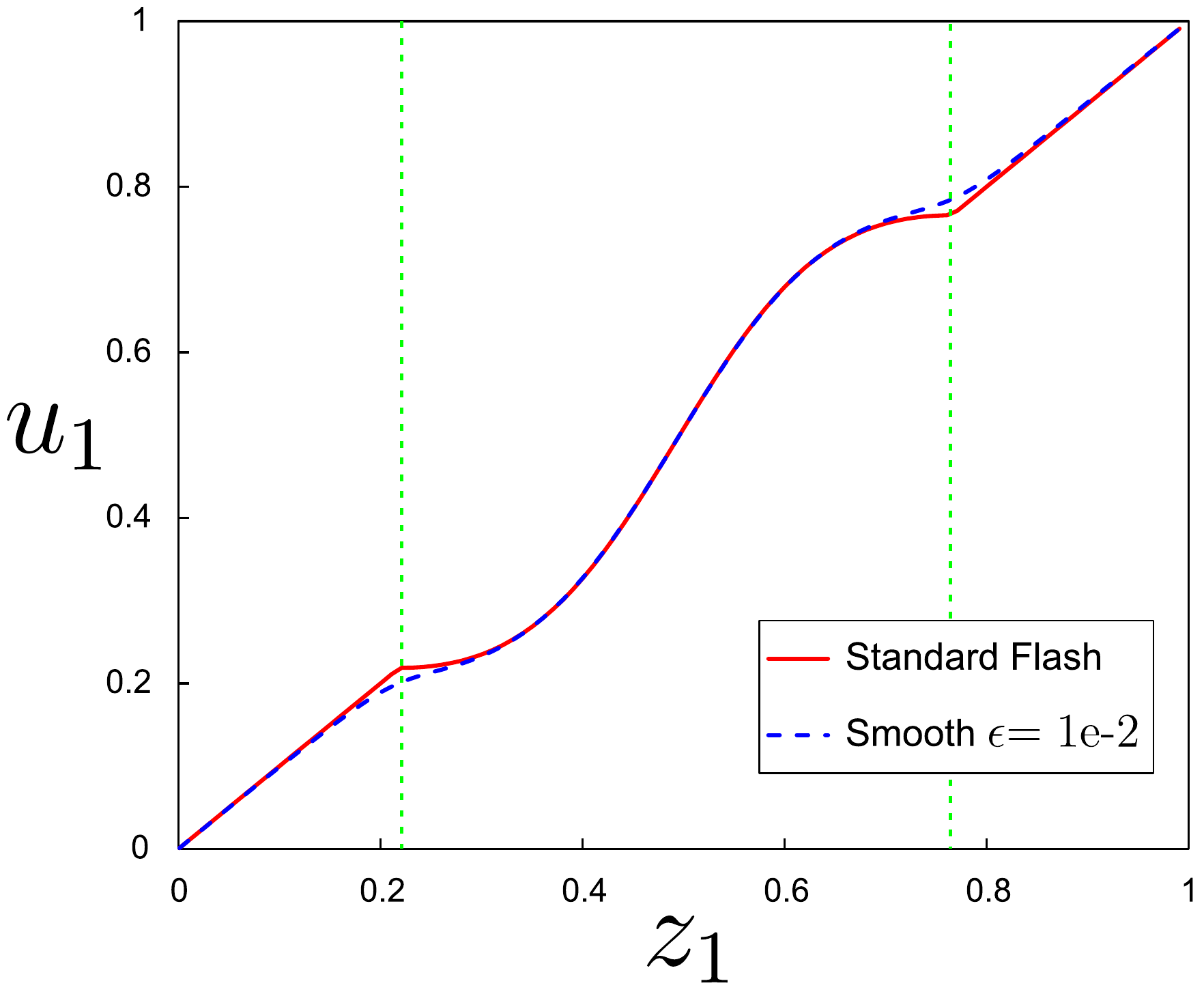}}
\caption{$\nu_g$ and $u_1$ of the two-component fluid under the different flash formulations.}
\label{fig:2comp_g_flux}
\end{figure}

Again we consider the two-component system from the last section.~We compare the solutions between the standard flash and smooth formulations.~\textbf{Fig.~\ref{fig:2comp_g_o_comp}} shows the compositions $y_1$ and $x_1$ as a function of $z_1$.~\textbf{Fig.~\ref{fig:2comp_g_flux}} compares the results of $\nu_g$ and $u_1$.

Note that the negative flash concept (Whitson and Michelsen 1989) is applied for allowing phase fractions (saturations) to exceed the bounds of 0 and 1 in the standard flash. Even though the compositions are constant along the tie-line, the discontinuities across the phase boundaries will still appear in the accumulation and flux terms.~In contrast, we can see that the smooth formulation with a fixed $\epsilon$ provides continuously differentiable functions of all the variables.

%==========================================================================================

\begin{figure}[!htb]
\centering
\subfloat[Composition $y_1$ ]{
\includegraphics[scale=0.5]{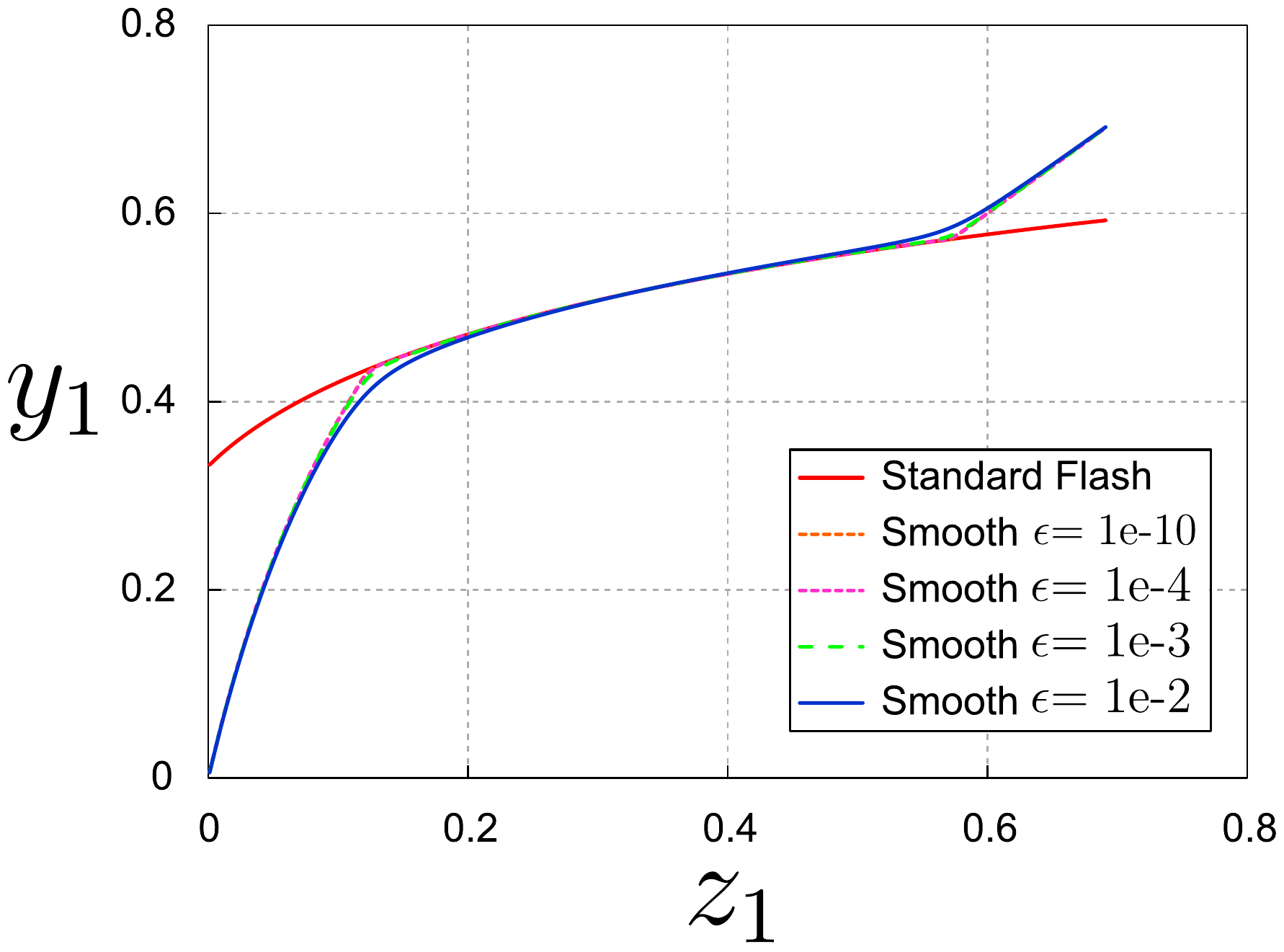}}
\\
\subfloat[Composition $x_1$ ]{
\includegraphics[scale=0.5]{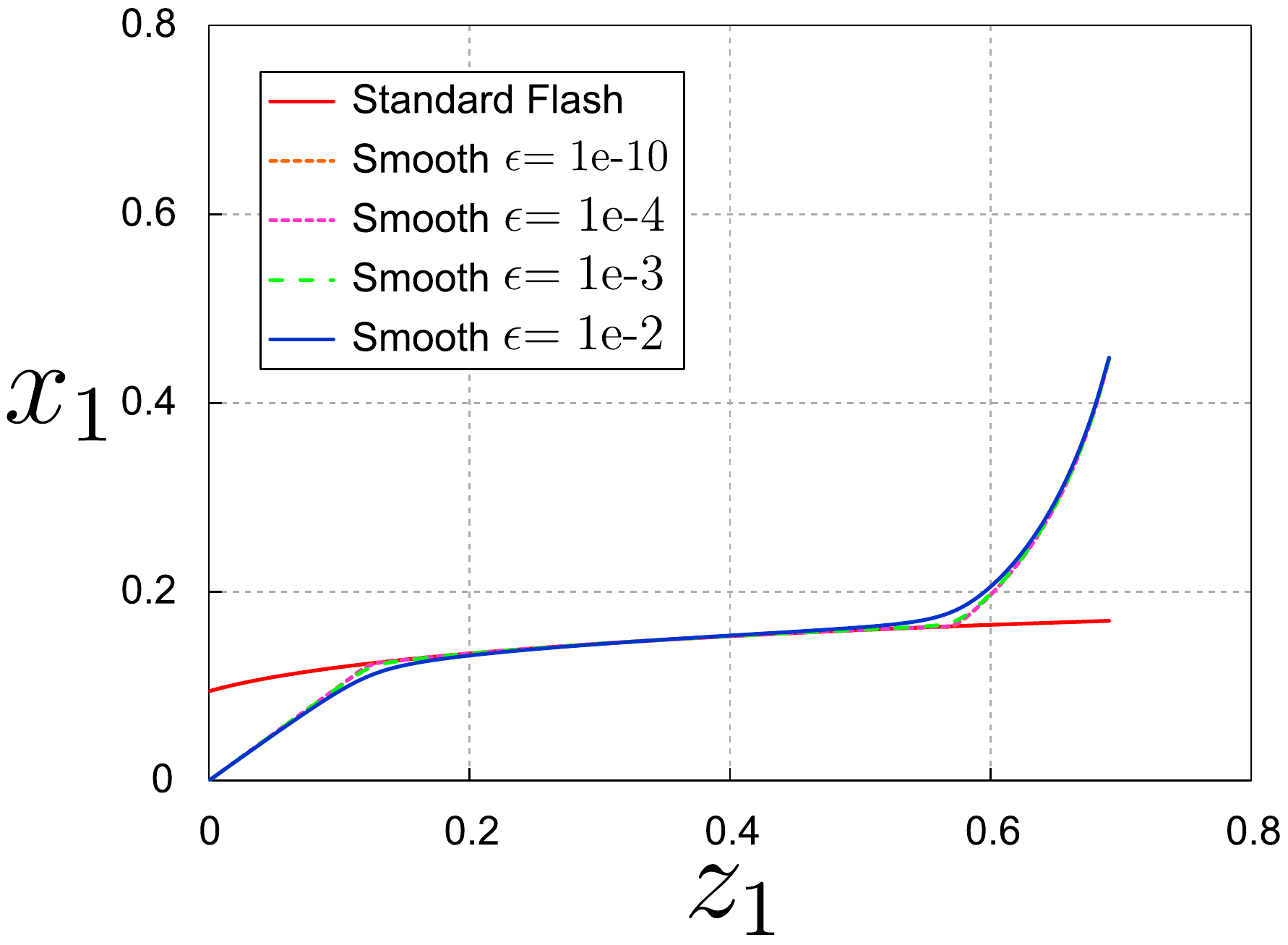}}
\caption{Compositions of the three-component fluid under the different flash formulations.}
\label{fig:3comp_g_o_comp}
\end{figure}

We also study a ternary system with $\left \{ K_1 = 3.5 , K_2 = 1.5, K_3 = 0.2 \right \}$ and quadratic relative permeability. The composition $z_2 = 0.3$ has constant value. \textbf{Fig.~\ref{fig:3comp_g_o_comp}} shows the compositions $y_1$ and $x_1$ as a function of $z_1$. We do not further plot overall fractional flow here because of its similar form as the two-component case. 

%==========================================================================================

\begin{figure}[!htb]
\centering
\subfloat[Composition $y_1$ ]{
\includegraphics[scale=0.5]{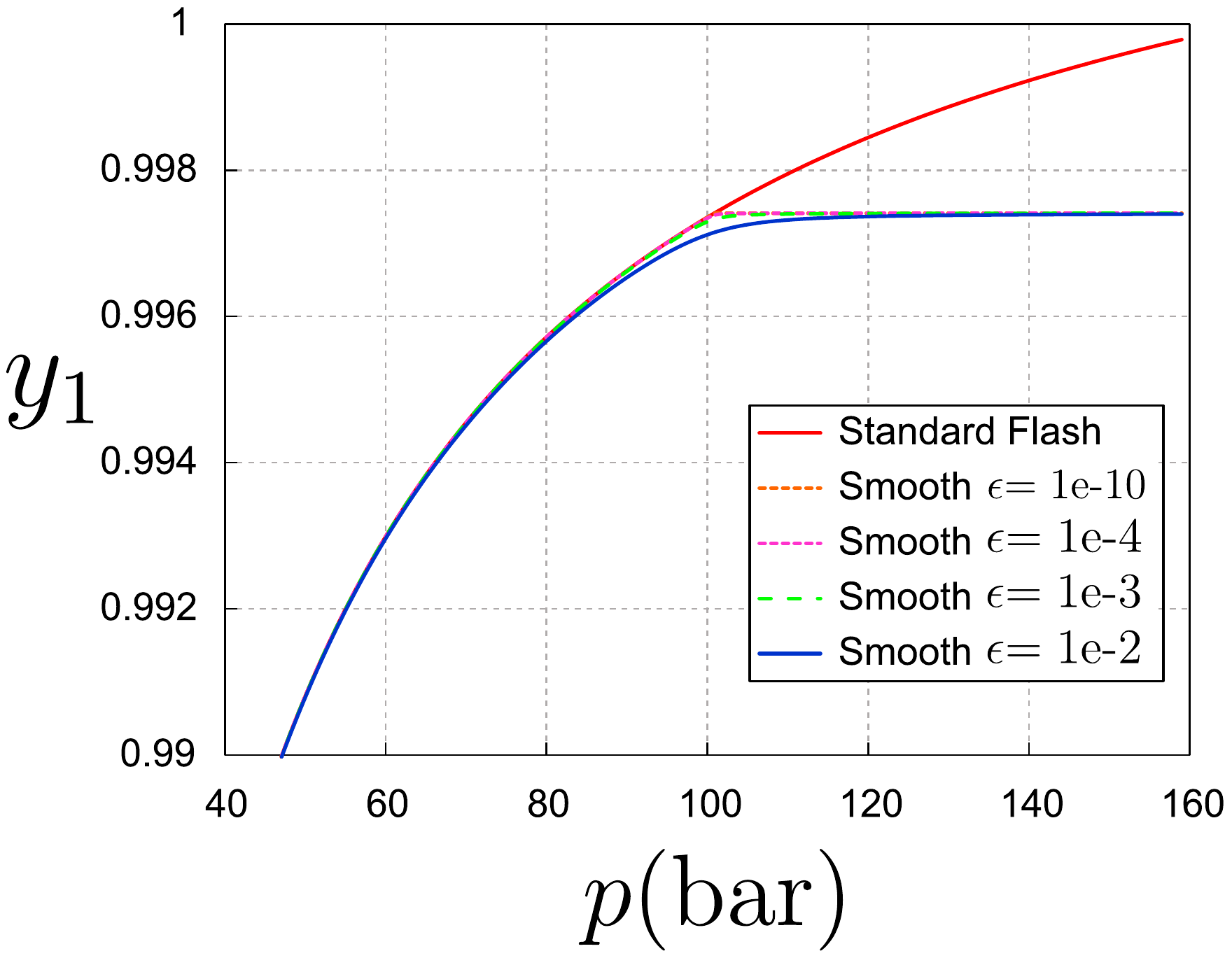}}
\\
\subfloat[Composition $x_1$ ]{
\includegraphics[scale=0.5]{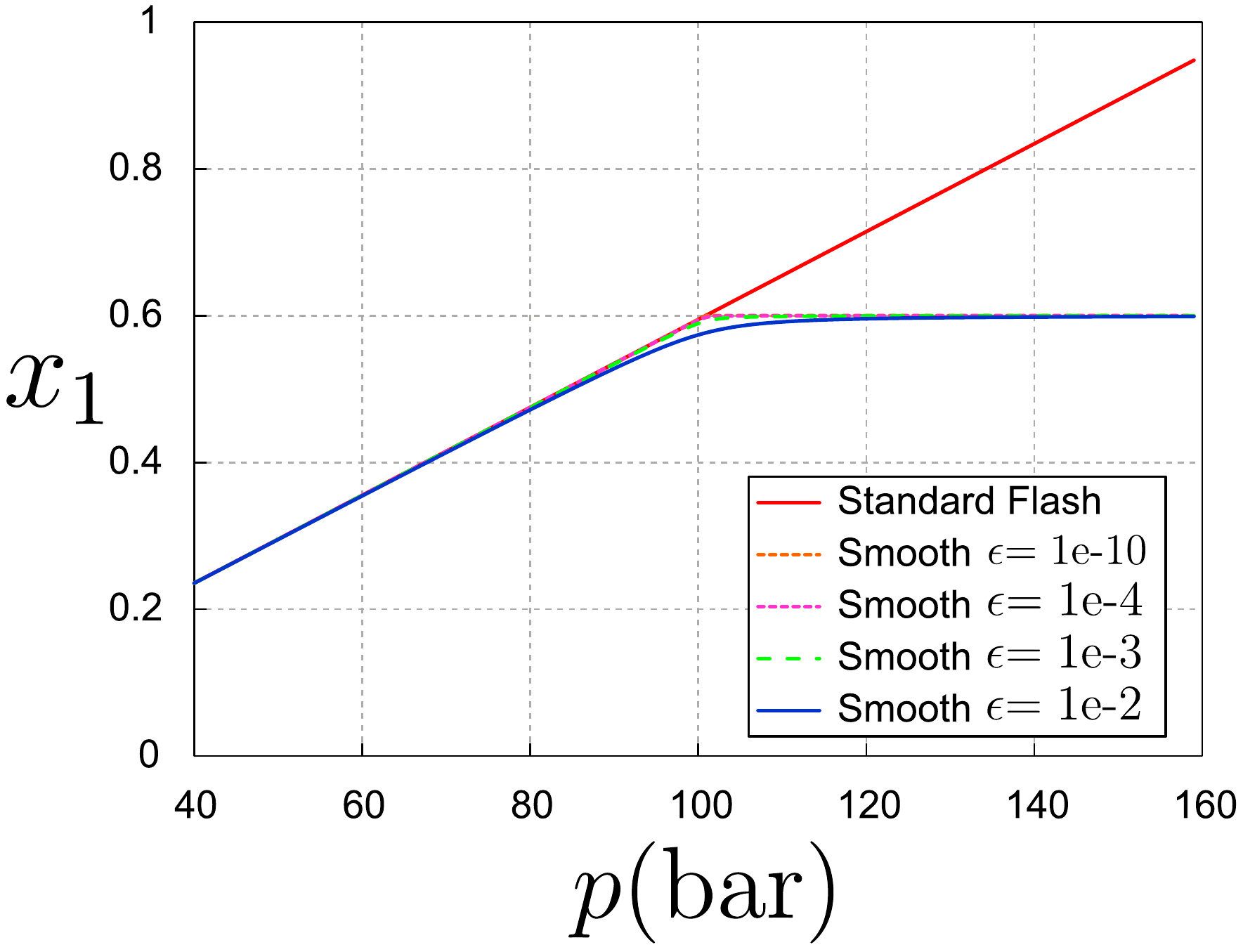}}
\caption{Compositions versus pressure under the different flash formulations.}
\label{fig:pres_g_o_comp}
\end{figure}

\begin{figure}[!htb]
\centering
\includegraphics[scale=0.5]{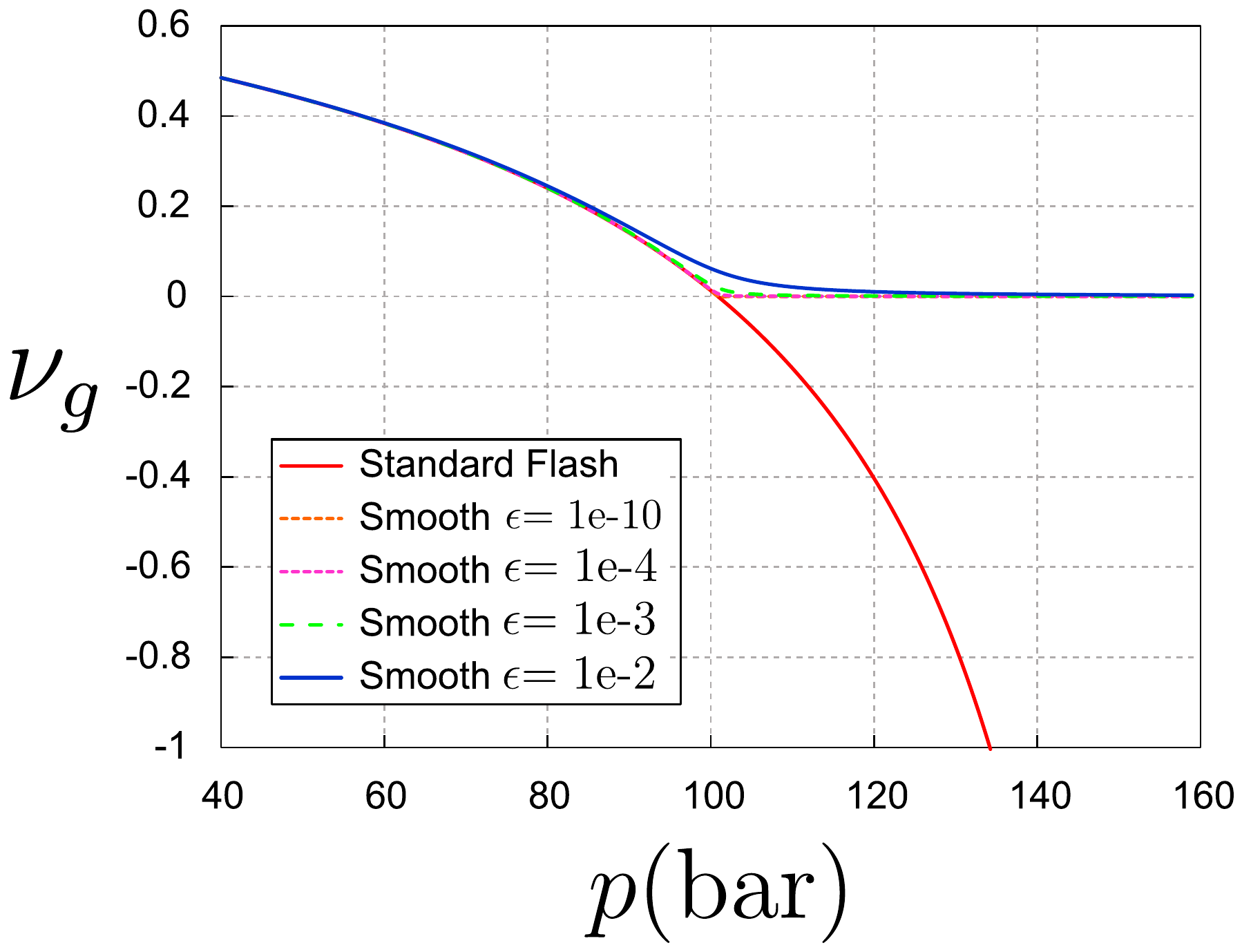}
\caption{Gas fraction versus pressure under the different flash formulations.}
\label{fig:pres_gas}
\end{figure}

In addition, we study a two-component fluid $\left \{ \textrm{C}_1 (60 \%), \textrm{C}_4 (40 \%) \right \}$ at a temperature of 250 $\textrm{K}$. The $K$-values for the two components are computed from Wilson's equation,  
\begin{equation} 
\label{eq:Kc_wil_eq}
K_c = \frac{\overline{p}_{c}}{p} \ \textrm{exp}\Bigg ( 5.37(1+\omega_c)\left ( 1- \frac{\overline{T}_{c}}{T} \right ) \Bigg )
\end{equation}
where $\overline{p}_{c}$ and $\overline{T}_{c}$ are critical pressure and temperature, respectively. We plot the compositions $y_1$ and $x_1$ as a function of $p$ in \textbf{Fig. \ref{fig:pres_g_o_comp}}. The gas phase fraction versus pressure is shown in \textbf{Fig. \ref{fig:pres_gas}}. Compared to the standard flash, the smooth formulation can achieve smooth transitions with respect to pressure. This is very beneficial for the compositional scenario driven by pressure.

%==========================================================================================

Lauser et al.~(2011) proposed a compositional formulation that is also based on complementarity conditions for handling phase changes. Compared to the smooth formulation proposed in this work, their approach has several limitations: 

1.~For a nonexistent phase in single-phase regimes, the sum of the molar fractions becomes less than one.~This can produce additional discontinuities in evaluating fluid properties.~Moreover, there is no guarantee that the values of molar fractions remain positive under all conditions, especially if smoothing approximations are applied.

2. Phase equilibrium and compositional systems are non-smooth in nature. Therefore a Newton solver will still suffer from the discontinuous issues due to phase changes.

3. The simulation studies demonstrated only insignificant improvement in nonlinear convergence, compared with the conventional natural formulation (Lauser et al.~2011; Gharbia et al. 2015). We give detailed descriptions and discussions on the approach of Lauser et al. (2011) in Appendix A. 

%==========================================================================================

\subsection{Smooth formulation for coupled system}

For the smooth compositional formulation, the primary unknowns include:

(1) $p$ $-$ pressure [1], 

(2) $s_l$ $-$ phase saturations [2],

(3) $\beta$ $-$ non-physical variable [1],

(4) $x_{c}$, $y_{c}$ $-$ phase compositions of each component [2$n_c$].
\\

The coupled system contains: the [$n_c$] conservation equations (\ref{eq:mass_con_comp}), the phase (\ref{eq:phase_const}) and saturation (\ref{eq:satu_const}) constraints, and the [$n_c$] relaxed equilibrium equations,
\begin{equation} 
y_c - \beta K_c x_c = 0 ,
\end{equation}
and the smoothing equation derived from (\ref{eq:mcp_smooth}), 
\begin{equation} 
\label{eq:mcp_smooth_so}
\Theta_{\epsilon} \Big (\epsilon, \left( 1 - s_o \right), \left ( \beta - 1 \right ), \left ( - s_o \right ) \Big ) = 0 .
\end{equation}
The smoothing parameter $\epsilon$ can be kept constant during simulation.~It is expected that a suitable value of $\epsilon$ will be chosen for a target class of problems.~Compared to the natural-variables set, an obvious advantage of the new smooth formulation is that the equations and unknowns are the same for any phase state.~The complex variable-substitution process is thus avoided.

For general multi-component models, phase boundaries are quite complex, which makes it difficult and costly to apply damping (safeguarding) based solution strategies. One particular type of discontinuity is the transition between two tie-lines (Orr 2007).

As previously demonstrated, the saturations are bounded to [0, 1] in the smooth system, so that the physical limiting is not necessary for the conservation equations. As a result, all the discontinuities associated with the phase changes transfer to the single and concise equation (\ref{eq:mcp_smooth_so}), and get removed through the smoothing approximation. The coupled system becomes smooth across the entire phase boundaries, bringing enormous benefits for Newton solvers.

Here we mainly focus on the $K$-value method to describe the phase equilibrium. Our first application target is on black-oil models (with dissolved-gas and vaporized-oil data). The developed smooth formulation can be readily applied to general EoS-based problems. Some discussions and preliminary test results are given in Appendix C.

\section{Nonlinear solver}

The spatial and temporal discretization schemes used for the compositional flow model are summarized in Appendix B.

\subsection{Newton method}

At each timestep of a FIM simulation, given the unknown vector $U^n$ and a fixed timestep size $\Delta t$, we intend to obtain the new state $U^{n+1}$. 

The nonlinear system is cast in residual form and solved by the Newton method, 
\begin{equation} 
\mathcal{R}(U^{n+1}) = 0 ,
\end{equation}

The Newton method comprises a sequence of iterations, each involving the construction of a Jacobian matrix and solution of the resulting linear system,
\begin{equation} 
\mathcal{J}(U^{\eta}) \Delta U^{\eta + 1} = - \mathcal{R}(U^{\eta}) ,
\end{equation}
where
\begin{equation} 
\Delta U^{\eta + 1} = U^{\eta +1} - U^{\eta} .
\end{equation}
and $\mathcal{J}(U^{\eta}) = \left. \frac{\partial \mathcal{R}}{\partial U} \right|_{U^\eta }$ denotes the Jacobian matrix of $\mathcal{R}$ with respect to $U^{\eta}$. The iterative process is performed until the nonlinear system is converged.

The formulation and method developed in this work are integrated into the Automatic Differentiation General Purpose Research Simulator (AD-GPRS) (Younis and Aziz 2007; Zhou et al. 2011; Voskov and Tchelepi 2012). The nonlinear framework of AD-GPRS is built on top of Automatic Differentiation with Expression Templates Library (ADETL). The Jacobian is automatically assembled and analytically derived through ADETL.

\subsection{Homotopy continuation method}

%   An update schedule of ... is specified in the DBC algorithm to vary the dissipation level.

The Newton process may often converge slowly, or even diverge, due to poor initial guess and large timestep size (Younis et al. 2010). The smooth formulation has nice global convergence property. However, solution accuracy may be degraded, with a fixed smoothing parameter $\epsilon$. The value of $\epsilon$ needs to be adaptively determined for an optimal balance between accuracy and nonlinear performance. 

The coupled system with the smoothing equation (\ref{eq:mcp_smooth_so}) can be viewed as a homotopy mapping $\mathcal{H}$. The objective is to solve the original system $\mathcal{R}(U) = 0$ containing the non-smooth mid function. We can see that $\mathcal{H}(U,0 ) = \mathcal{R}(U)$ with the modified residual $\mathcal{H}(U, \epsilon)$ and the smoothing (continuation) parameter $\epsilon \geq 0$. Consider that $\mathcal{H}(U,\epsilon_0) = 0$ is much easier to solve than the target problem $\mathcal{H}(U,0 ) = 0$. The continuation method can be developed by discretizing in $\epsilon$ to form a sequence of nonlinear systems $\mathcal{H}(U, \epsilon_{\eta}) = 0$. The target solution is reached by progressively decreasing $\epsilon$ from $\epsilon_0$ towards 0 to globalize the flow equations. As a result, the algorithm will not have an impact on the accuracy of the final solutions. 

The continuation method only acts as a globalization stage to obtain better initial guesses for the Newton process (Jiang and Tchelepi 2018). In our experience, applying the continuation for a few iterations in the globalization stage is already highly effective. A simple strategy is used to evolve $\epsilon$ during each timestep, 
\begin{equation} 
\label{eq:sbc_form}
\epsilon_{\eta} = \begin{cases}
\textrm{max} \left ( \gamma \epsilon_{\eta-1} \ , \ \epsilon_{min} \right ) \ , & \ \eta \leq \eta_{max} \\ 
\epsilon_{min} \ , & \ \eta > \eta_{max}
\end{cases}
\end{equation}
with $\epsilon_0$ as the initial value of a timestep. After each Newton iteration, $\epsilon$ is multiplied by a constant $\gamma$. $\eta_{max}$ is the number of iterations taken for globalization. The parameter values used for all the following cases are: $\epsilon_0 = 0.1$, $\eta_{max} = 4$ and $\gamma = 0.5$. $\epsilon_{min}$ is specified to ensure that the system is smooth enough at the convergence limit. From the simulations we find that a value around $\epsilon_{min} = \textrm{1.0e-4}$ can produce the solutions with satisfactory accuracy and nonlinear convergence. 

We note that the smooth formulation and the continuation method involve different complexities and efforts for implementation. Subsequently, their applications depend on specific accuracy and implementation considerations.

\section{Results: 1D model}

% unless otherwise stated
% The simulation example presented previously is still employed in this section. 

%----------------------------------------------------------------------------------

Nonlinear convergence is based on the following criterion: $\textrm{max}_{i,c}\left | R_{i,c} \right | < 10^{-4}$, where $R_{i,c}$ is the residual of $c$-th equation in $i$-th cell (for conservation equations, we normalize by the total mass of $c$-th component).

A simple time-stepping strategy is employed:~if the Newton method converges for the current timestep, the next timestep will be doubled; if the nonlinear solver fails to converge, the timestep is reduced by half and solved again.~The solution from previous timestep $n$ is taken to be the initial guess for a new timestep.~During the iterative process, all fractions in the variable-set are kept within the physical interval [0, 1].~Also the damping strategy is employed to stabilize Newton updates: the local chopping (maximum allowable change) values of 0.2 are used for saturations and 0.1 for molar fractions.

We evaluate the efficiency of the new approaches using several complex problems. Five different fluids, and two different reservoir models are considered. The models include: (a) homogeneous 1D model; (b) heterogeneous 2D model taken from the bottom layer of the SPE 10 problem. In the following cases, simple relative permeabilities given by quadratic function are used, unless otherwise indicated. Note that both phase density and viscosity depend on pressure and compositions. Phase molar density $\rho_l$ is evaluated based on the compressibility (Z) factor from the Peng-Robinson EoS. Phase viscosity $\mu_l$ is computed by the correlation of Lohrenz et al. (1964). The specification of the 1D base model is provided in Table \ref{tab:specification_m}.

\begin{table}[!htb]
\centering
\caption{Specification of the 1D base model}
\label{tab:specification_m}
\begin{tabular}{|c|c|c|}
\hline
Parameter                  &  Value            & Unit   \\ \hline
NB                         &  500              &        \\ \hline
DX / DY / DZ               &  10 / 10 / 10     & m      \\ \hline
Permeability               &  1000             & md     \\ \hline
Porosity                   &  0.2              &        \\ \hline
Rock compressibility       &  1e-5             & 1/bar  \\ \hline

Max timestep size          &  50               & day    \\ \hline
Total simulation time      &  500              & day    \\ \hline
Max number of nonlinear iterations     & 20         &     \\ \hline
\end{tabular}
\end{table}

\subsection{Two-component fluid}

We test a comprehensive suite of 1D problems. The first problem is the displacement of propane $\textrm{C}_3$ by methane $\textrm{C}_1$ in a horizontal domain. Cell size of 1m is used. The fluid and compositions are given in Table \ref{tab:2comp_fluid}. $Z_{i}$ and $Z_{inj}$ are initial and injection compositions, respectively. Pressure is kept constant as 65 bars at the production end. A constant volumetric rate 5 $\textrm{m}^3/\textrm{day}$ is specified at the injection end. Initial pressure is 70 bars and temperature is 311 $\textrm{K}$.

\begin{table}[!htb]
\centering
\caption{Fluid and compositions of the two-component case.}
\label{tab:2comp_fluid}
\begin{tabular}{|c|c|c|c|}
\hline
Comp    & $K$   & $Z_{i}$    & $Z_{inj}$   \\ \hline
$C_1$   & 3.5   & 0.001      & 0.99        \\ \hline
$C_3$   & 0.5   & 0.999      & 0.01        \\ \hline
\end{tabular}
\end{table}

\subsubsection{Case 1}

The nonlinear performance of the two-component case is summarized in Table \ref{tab:case_1d_2c_1}. We report the total number of Newton iterations and timesteps. In parentheses, we also give the number of wasted iterations that correspond to the iterations spent on unconverged timesteps. `SBC' denotes the smoothing based continuation method, and $\epsilon_{min}$ is specified for Eq. (\ref{eq:sbc_form}). For each simulation, the maximum CFL number, averaged over the timesteps taken, is reported. The corresponding maxCFL of the case is 16.

As we can see from the results, there is no timestep cut during the simulations. Even for this simple scenario, the SBC method requires much fewer iterations to converge.

\begin{table}[!htb]
\centering
\caption{Computational performance of Case 1 with the two-component fluid.}
\label{tab:case_1d_2c_1}
\begin{tabular}{|c|c|c|}
\hline
                                                           & Total iterations (Wasted) & Timesteps (Wasted) \\ \hline
\begin{tabular}[c]{@{}c@{}}Standard\\ Natural\end{tabular} & 188 (0)                   & 12 (0)             \\ \hline
\begin{tabular}[c]{@{}c@{}}SBC\\ $\epsilon_{min} = \textrm{1.0e-4}$\end{tabular}            & 115 (0)                   & 12 (0)             \\ \hline
\begin{tabular}[c]{@{}c@{}}SBC\\ $\epsilon_{min} = \textrm{1.0e-2}$\end{tabular}            & 107 (0)                   & 12 (0)             \\ \hline
\end{tabular}
\end{table}

%==========================================================

\begin{figure}[!htb]
\centering
\subfloat[Gas saturation]{
\includegraphics[scale=0.6]{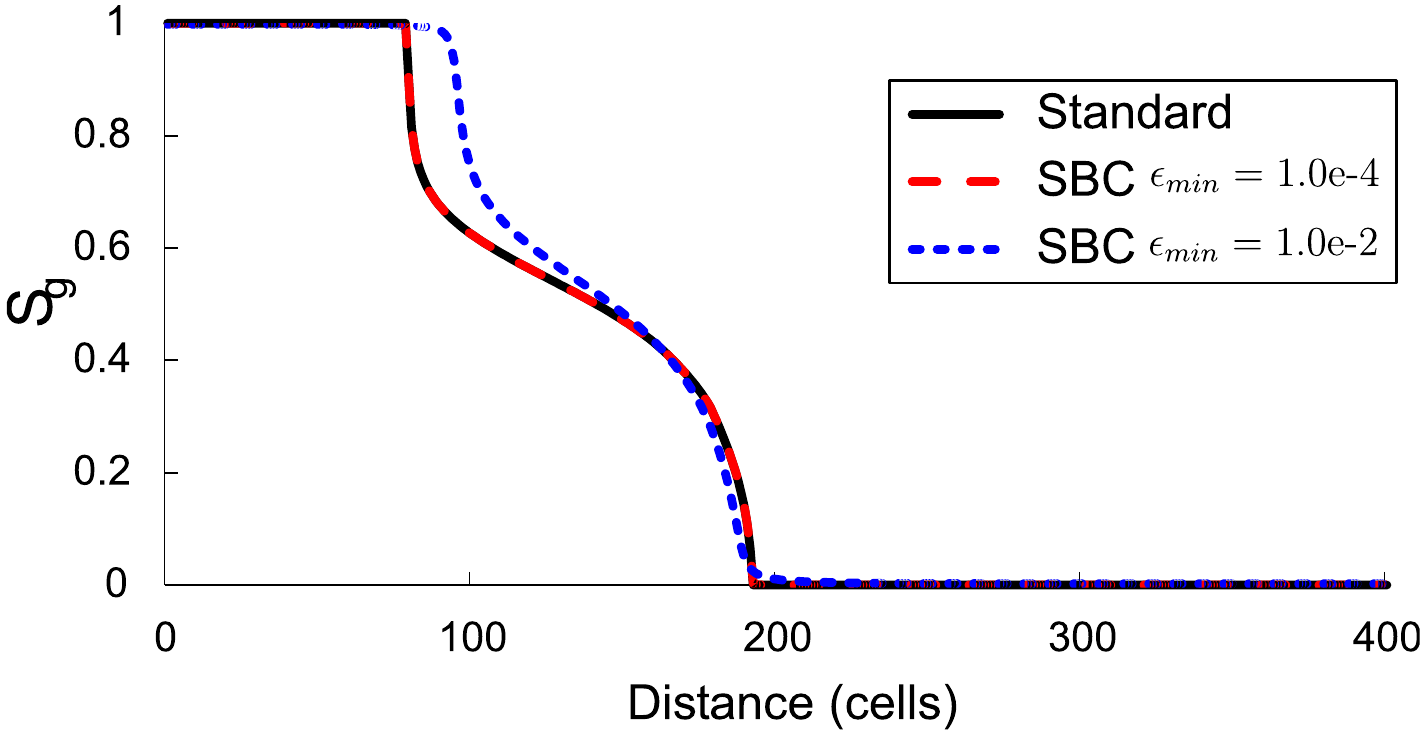}}
\\
\subfloat[Overall composition of $C_1$]{
\includegraphics[scale=0.6]{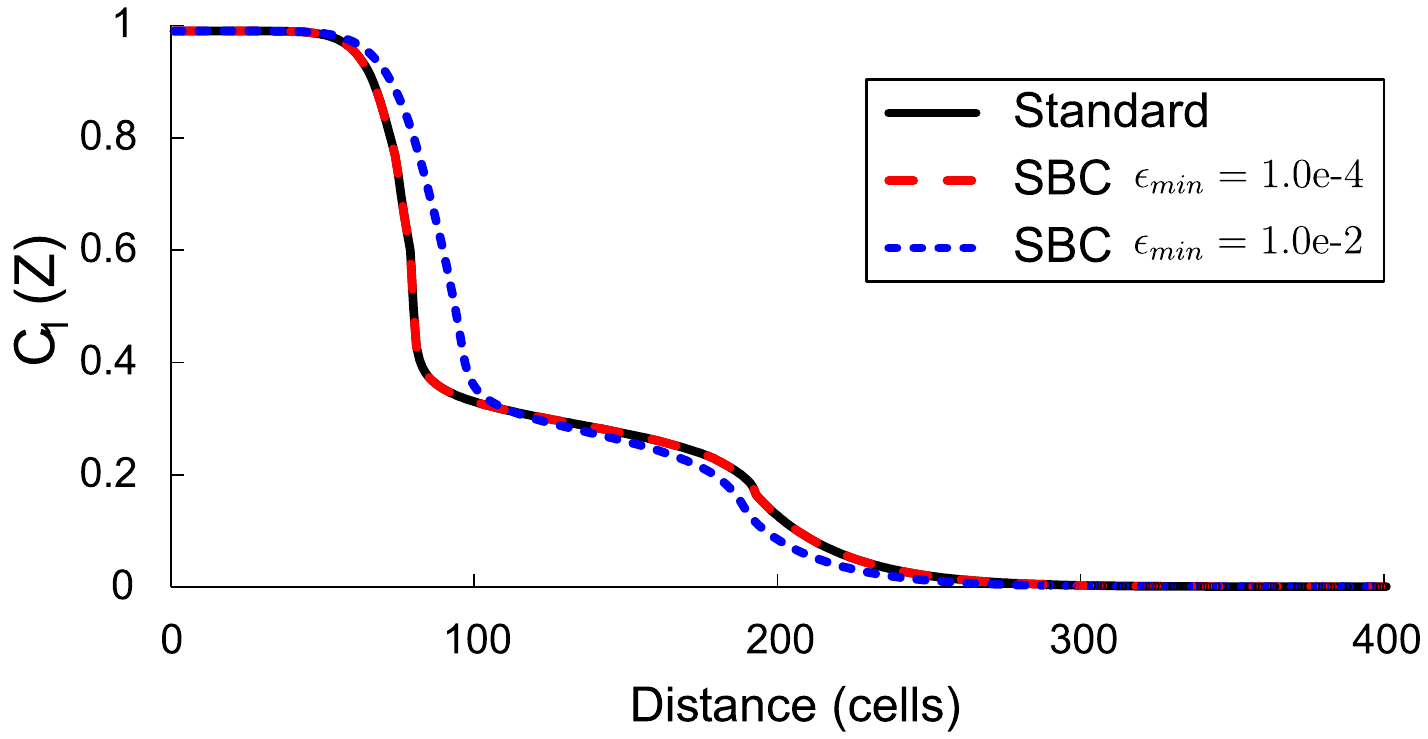}}
\caption{Gas saturation and overall composition profiles of Case 1 with the two-component fluid.}
\label{fig:2comp_1}
\end{figure}

Gas saturation and overall composition profiles are plotted in \textbf{Fig. \ref{fig:2comp_1}}. The profiles show two shocks formed in the two-phase region. The solutions from SBC with $\epsilon_{min} = \textrm{1.0e-4}$ fully match the standard natural formulation. For $\epsilon_{min} = \textrm{1.0e-2}$, some differences around the shocks are observed.

\subsubsection{Case 2}

We also test a case with gravity. Temperature is set to 380 $\textrm{K}$. The maximum time step size reduces to 30 days, due to severe degradation of nonlinear convergence encountered by the standard natural formulation. The total simulation time becomes 400 days. The other parameters from the previous case remain unchanged. 

The nonlinear performance of the case with gravity is summarized in Table \ref{tab:case_1d_2c_2}. The maxCFL is 20. As can be seen, the standard formulation suffers from multiple timestep cuts and subsequent wasted iterations. In contrast, the SBC method does not require any timestep cut, resulting in much smaller number of iterations.

\begin{table}[!htb]
\centering
\caption{Computational performance of the two-component case with gravity.}
\label{tab:case_1d_2c_2}
\begin{tabular}{|c|c|c|}
\hline
                                                           & Total iterations (Wasted) & Timesteps (Wasted) \\ \hline
\begin{tabular}[c]{@{}c@{}}Standard\\ Natural\end{tabular} & 430 (140)                 & 28 (7)             \\ \hline
\begin{tabular}[c]{@{}c@{}}SBC\\ $\epsilon_{min} = \textrm{1.0e-4}$\end{tabular}            & 201 (0)                   & 18 (0)             \\ \hline
\begin{tabular}[c]{@{}c@{}}SBC\\ $\epsilon_{min} = \textrm{1.0e-2}$\end{tabular}            & 176 (0)                   & 18 (0)             \\ \hline
\end{tabular}
\end{table}

\subsubsection{Case 3}

We further study a case with variable $K$-values computed from Eq.~(\ref{eq:Kc_wil_eq}). Production pressure changes to 70 bars. Initial pressure is 80 bars at a temperature of 370 $\textrm{K}$. The maximum timestep size becomes 50 days, with a total simulation time of 400 days.

Here we simply apply the smooth formulation without the continuation method. The case is to demonstrate the applicability of the smooth formulation using a fixed smoothing parameter $\epsilon$. The value level of $\epsilon = \textrm{1.0e-2}$ is sufficient to provide much improved convergence performance. Solution profiles are not shown because the differences between the two formulations are small.

The nonlinear performance of Case 3 is summarized in Table \ref{tab:case_1d_2c_3}. The maxCFL is 21. As can be seen, the pressure-dependent $K$-values make it very challenging for the standard formulation. By comparison, the smooth formulation is continuously differentiable with respect to pressure, leading to significant convergence speedup. 

%==================================================================

\begin{table}[!htb]
\centering
\caption{Computational performance of Case 3 with the two-component fluid.}
\label{tab:case_1d_2c_3}
\begin{tabular}{|c|c|c|}
\hline
                                                           & Total iterations (Wasted) & Timesteps (Wasted) \\ \hline
\begin{tabular}[c]{@{}c@{}}Standard\\ Natural\end{tabular} & 516 (260)                   & 33 (13)             \\ \hline
\begin{tabular}[c]{@{}c@{}}Smooth \end{tabular}            & 110 (0)                     & 13 (0)             \\ \hline
\end{tabular}
\end{table}

%==================================================================

\subsection{Three-component fluid}

We test cases with three-component fluid systems. Cubic relative permeabilities are used. Initial pressure is 70 bars and temperature is 320 $\textrm{K}$. Pressure is kept constant at the both injection and production ends. Production pressure is 60 bars, with injection pressure as 130 bars. The maximum timestep size is 50 days, with a total simulation time of 400 days.

\subsubsection{Case 1}

The fluid and compositions for Case 1 are given in Table \ref{tab:3comp_fluid_1}.

%==================================================================

\begin{table}[!htb]
\centering
\caption{Fluid and compositions of the three-component case.}
\label{tab:3comp_fluid_1}
\begin{tabular}{|c|c|c|c|}
\hline
Comp      & $K$   & $Z_{i}$   & $Z_{inj}$  \\ \hline
$C_1$     & 2.5   & 0.01      & 0.97       \\ \hline
$C_4$     & 0.6   & 0.5       & 0.02       \\ \hline
$C_{10}$  & 0.2   & 0.49      & 0.01       \\ \hline
\end{tabular}
\end{table}

%==================================================================

The nonlinear performance of Case 1 is summarized in Table \ref{tab:case_1d_3c_1}. The maxCFL is 6. In this case, the SBC method does not require any timestep cuts.

\begin{table}[!htb]
\centering
\caption{Computational performance of Case 1 with the three-component fluid.}
\label{tab:case_1d_3c_1}
\begin{tabular}{|c|c|c|}
\hline
                                                           & Total iterations (Wasted) & Timesteps (Wasted) \\ \hline
\begin{tabular}[c]{@{}c@{}}Standard\\ Natural\end{tabular} & 320 (100)                 & 27 (5)             \\ \hline
\begin{tabular}[c]{@{}c@{}}SBC\\ $\epsilon_{min} = \textrm{1.0e-4}$\end{tabular}            & 175 (0)                   & 21 (0)             \\ \hline
\begin{tabular}[c]{@{}c@{}}SBC\\ $\epsilon_{min} = \textrm{1.0e-2}$\end{tabular}            & 164 (0)                   & 21 (0)             \\ \hline
\end{tabular}
\end{table}

%==================================================================

\begin{figure}[!htb]
\centering
\subfloat[Gas saturation]{
\includegraphics[scale=0.6]{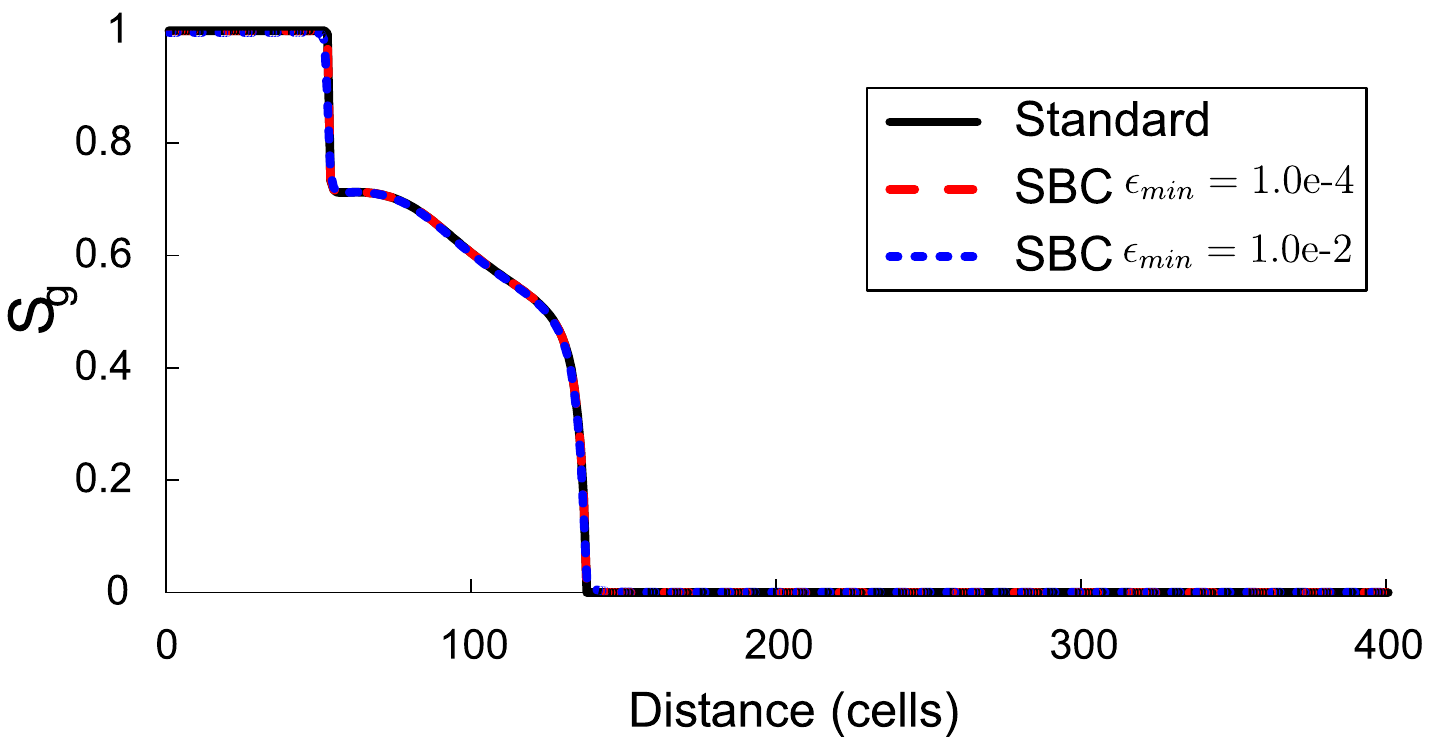}}
\\
\subfloat[Overall composition of $C_1$]{
\includegraphics[scale=0.6]{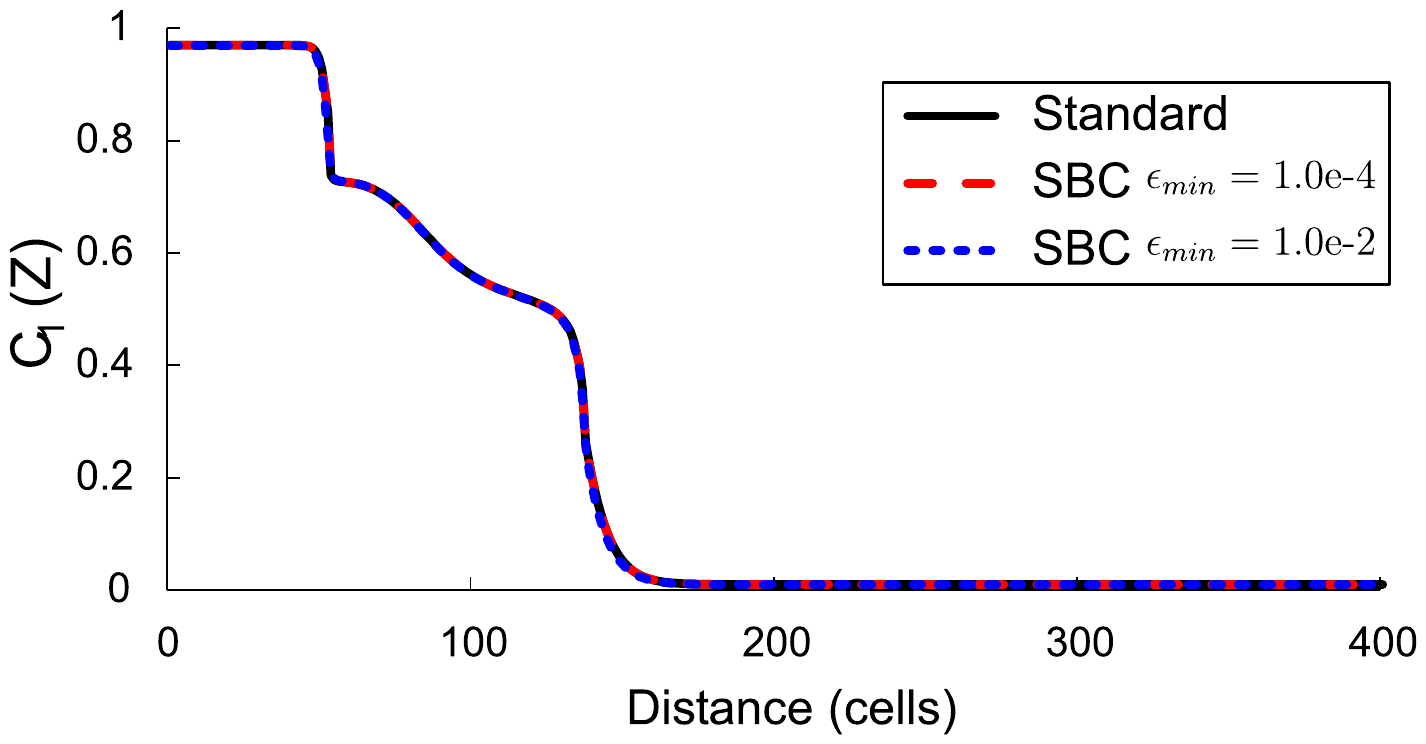}}
\caption{Gas saturation and overall composition profiles of Case 1 with the three-component fluid.}
\label{fig:3comp_1}
\end{figure}

The gas saturation and overall composition profiles are plotted in \textbf{Fig. \ref{fig:3comp_1}}. Here the timestep size is reduced to 10 days, to ensure that the solutions from the different methods are compared under the same time-stepping schedule. As we can see, SBC produces solutions that are very close to the standard formulation.

\subsubsection{Case 2}

We also study a three-component system with $\left \{ K_1 = 2.5 , K_4 = 1.5, K_{10} = 0.3 \right \}$, and the other parameters remain unchanged. 

The nonlinear performance of Case 2 is summarized in Table \ref{tab:case_1d_3c_2}. The maxCFL is 6. The standard formulation suffers from a large number of timestep cuts and wasted iterations, despite a small CFL number for this case. The SBC method effectively stabilizes the iterative process, leading to superior convergence performance.

\begin{table}[!htb]
\centering
\caption{Computational performance of Case 2 with the three-component fluid.}
\label{tab:case_1d_3c_2}
\begin{tabular}{|c|c|c|}
\hline
                                                           & Total iterations (Wasted) & Timesteps (Wasted) \\ \hline
\begin{tabular}[c]{@{}c@{}}Standard\\ Natural\end{tabular} & 605 (320)                 & 45 (16)             \\ \hline
\begin{tabular}[c]{@{}c@{}}SBC\\ $\epsilon_{min} = \textrm{1.0e-4}$\end{tabular}            & 167 (0)                   & 20 (0)             \\ \hline
\begin{tabular}[c]{@{}c@{}}SBC\\ $\epsilon_{min} = \textrm{1.0e-2}$\end{tabular}            & 144 (0)                   & 20 (0)             \\ \hline
\end{tabular}
\end{table}

\subsection{Case 3}

We consider a gas-condensate mixture with the fluid and compositions given in Table \ref{tab:3comp_fluid_3}. Quadratic relative permeabilities are used. Initial, production, injection pressures are 85, 80, 120 bars, respectively. Temperature is 325 $\textrm{K}$.

%==================================================================

\begin{table}[!htb]
\centering
\caption{Fluid and compositions of the gas-condensate system with three-component fluid.}
\label{tab:3comp_fluid_3}
\begin{tabular}{|c|c|c|c|}
\hline
Comp      & $K$    & $Z_{i}$  & $Z_{inj}$  \\ \hline
$C_1$     & 2.5    & 0.5      & 0.98       \\ \hline
$C_2$     & 1.5    & 0.4      & 0.01       \\ \hline
$C_{5}$   & 0.05   & 0.1      & 0.01       \\ \hline
\end{tabular}
\end{table}

%==================================================================

\begin{figure}[!htb]
\centering
\subfloat[Gas saturation]{
\includegraphics[scale=0.6]{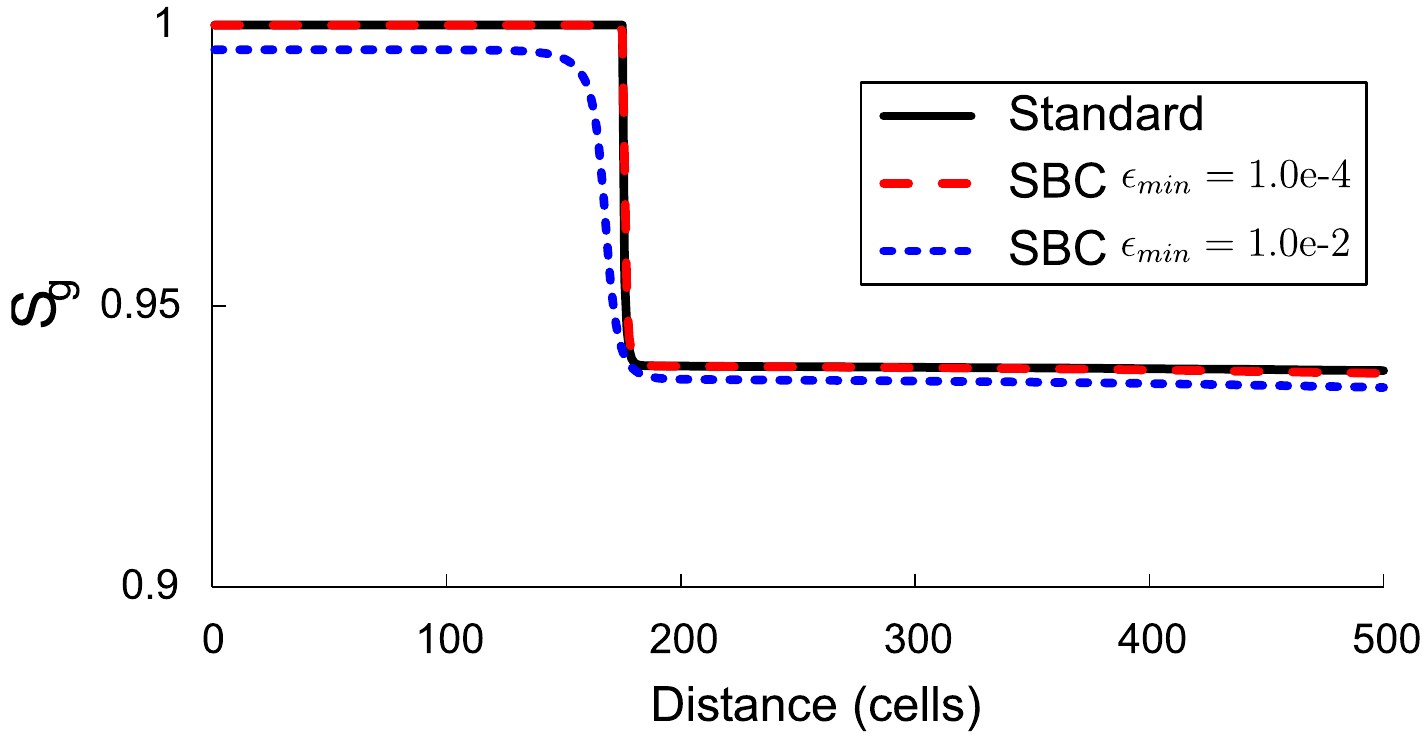}}
\\
\subfloat[Overall composition of $C_1$]{
\includegraphics[scale=0.6]{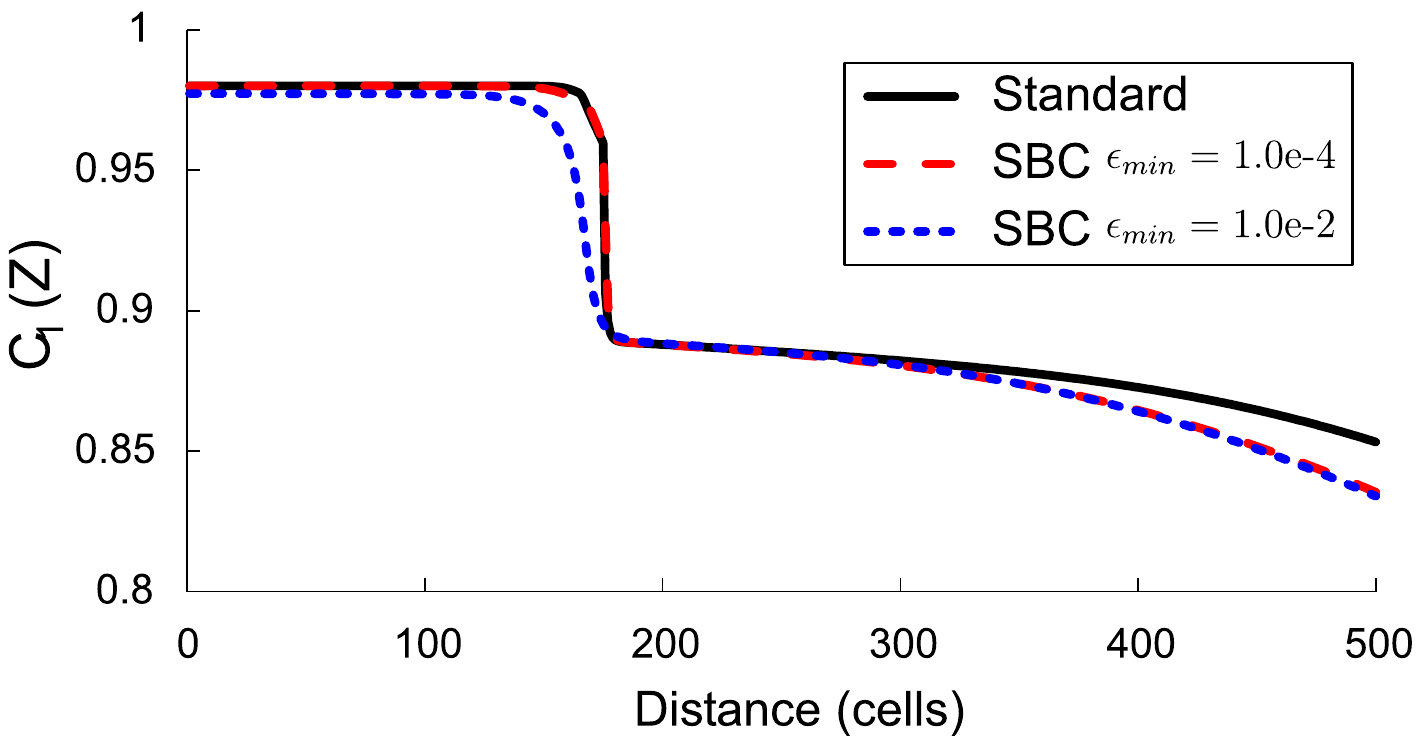}}
\caption{Gas saturation and overall composition profiles of Case 3 with the three-component fluid.}
\label{fig:3comp_3}
\end{figure}

The nonlinear performance of the gas-condensate case is summarized in Table \ref{tab:case_1d_3c_gc}. The maxCFL is 54. The gas saturation and overall composition profiles are plotted in \textbf{Fig. \ref{fig:3comp_3}}. The gas-condensate mixture forms a sharp front in the domain. For $\epsilon_{min} = \textrm{1.0e-2}$, small errors are produced around the shock. Note that the solution difference near the production end between SBC and the standard formulation is due to different time-stepping schedules. 

\begin{table}[!htb]
\centering
\caption{Computational performance of the gas-condensate system with three-component fluid.}
\label{tab:case_1d_3c_gc}
\begin{tabular}{|c|c|c|}
\hline
                                                           & Total iterations (Wasted) & Timesteps (Wasted) \\ \hline
\begin{tabular}[c]{@{}c@{}}Standard\\ Natural\end{tabular} & 413 (200)                 & 35 (10)             \\ \hline
\begin{tabular}[c]{@{}c@{}}SBC\\ $\epsilon_{min} = \textrm{1.0e-4}$\end{tabular}            & 150 (0)                   & 20 (0)             \\ \hline
\begin{tabular}[c]{@{}c@{}}SBC\\ $\epsilon_{min} = \textrm{1.0e-2}$\end{tabular}            & 143 (0)                   & 20 (0)             \\ \hline
\end{tabular}
\end{table}

\subsubsection{Case 4}

We further consider a case with variable $K$-values. Initial pressure is 100 bars, at a temperature of 340 $\textrm{K}$. Production and injection pressures are 95 and 140 bars, respectively. Initial compositions are $\left \{ C_1 (1 \%), C_2 (50 \%), C_{3} (49 \%) \right \}$, and injection mixture is $\left \{ C_1 (97 \%), C_2 (2 \%), C_{3} (1 \%) \right \}$. 

%==================================================================

The value of $\epsilon = \textrm{1.0e-2}$ is used for the smoothing parameter. The nonlinear performance of Case 4 is summarized in Table \ref{tab:case_1d_3c_4}. The maxCFL is 24. The standard formulation shows a poor nonlinear performance, while the iteration number is largely reduced under the smooth formulation. 

\begin{table}[!htb]
\centering
\caption{Computational performance of Case 4 with the three-component fluid.}
\label{tab:case_1d_3c_4}
\begin{tabular}{|c|c|c|}
\hline
                                                           & Total iterations (Wasted) & Timesteps (Wasted) \\ \hline
\begin{tabular}[c]{@{}c@{}}Standard\\ Natural\end{tabular} & 644 (340)                   & 47 (17)             \\ \hline
\begin{tabular}[c]{@{}c@{}}Smooth \end{tabular}            & 150 (0)                     & 20 (0)             \\ \hline
\end{tabular}
\end{table}

\subsection{Four-component fluid}

We use a four-component fluid system, comprised of $\left \{ C_1 , CO_2 , C_4 , C_{10} \right \}$ throughout our calculation examples.

\subsubsection{Case 1}

The fluid and compositions are given in Table \ref{tab:4comp_fluid_1}. Initial pressure is 75 bars, at a temperature of 410 $\textrm{K}$. Production pressure is 70 bars, with injection pressure as 140 bars. The total simulation time is 400 days. 

%==================================================================

\begin{table}[!htb]
\centering
\caption{Fluid and compositions of the four-component fluid.}
\label{tab:4comp_fluid_1}
\begin{tabular}{|c|c|c|c|}
\hline
Comp        & $K$    & $Z_{i}$   & $Z_{inj}$   \\ \hline
$C_1$       & 2.5    & 0.2       & 0.028       \\ \hline
$CO_2$      & 1.5    & 0.01      & 0.97        \\ \hline
$C_4$       & 0.5    & 0.29      & 0.001       \\ \hline
$C_{10}$    & 0.05   & 0.5       & 0.001       \\ \hline
\end{tabular}
\end{table}

%==================================================================

The nonlinear performance of Case 1 is summarized in Table \ref{tab:case_1d_4c_1}. The maxCFL is 17. The case is to validate the applicability of the smooth formulation without the continuation method. We can see that the value of $\epsilon = \textrm{1.0e-2}$ brings a reduction in the iteration number, though both the formulations perform well for this case. 

\begin{table}[!htb]
\centering
\caption{Computational performance of Case 1 with the four-component fluid.}
\label{tab:case_1d_4c_1}
\begin{tabular}{|c|c|c|}
\hline
                                                           & Total iterations (Wasted) & Timesteps (Wasted) \\ \hline
\begin{tabular}[c]{@{}c@{}}Standard\\ Natural\end{tabular} & 160 (0)                 & 20 (0)             \\ \hline
\begin{tabular}[c]{@{}c@{}}Smooth\\ $\epsilon = \textrm{1.0e-4}$\end{tabular}            & 160 (0)                   & 20 (0)             \\ \hline
\begin{tabular}[c]{@{}c@{}}Smooth\\ $\epsilon = \textrm{1.0e-2}$\end{tabular}            & 110 (0)                   & 20 (0)             \\ \hline
\end{tabular}
\end{table}

%-------------------------------------------------------------------------------

\begin{figure}[!htb]
\centering
\subfloat[Gas saturation]{
\includegraphics[scale=0.6]{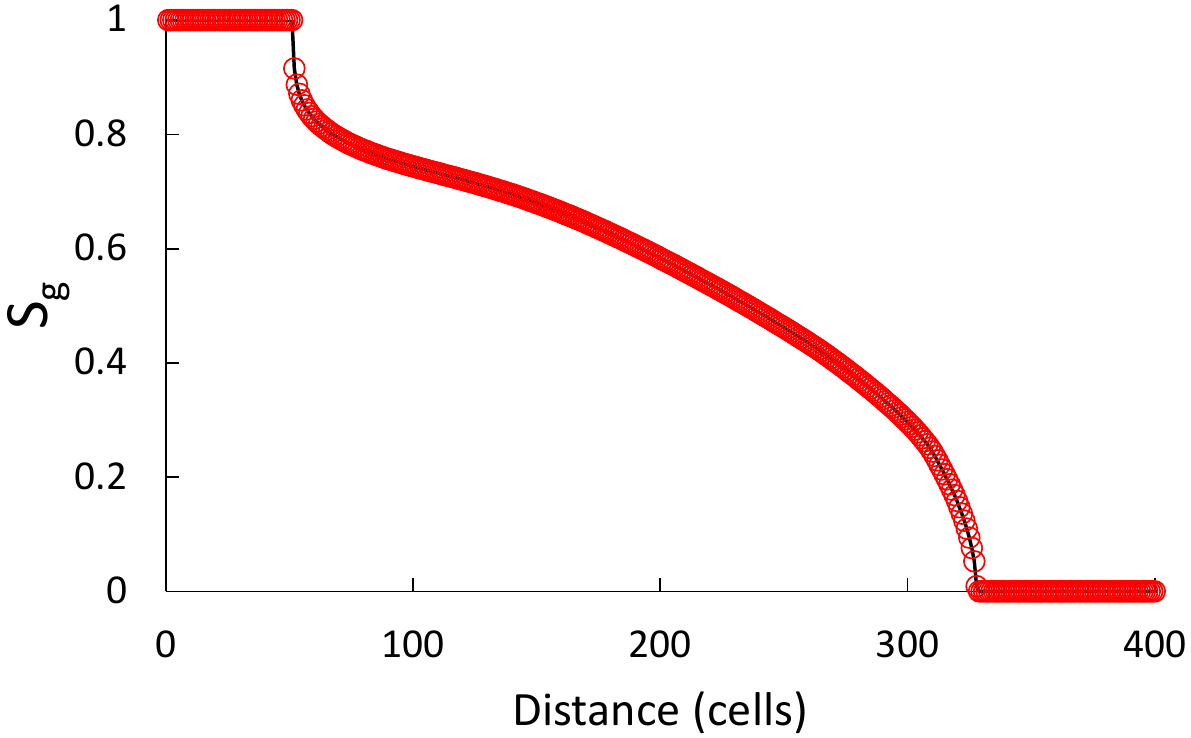}}
\\
\subfloat[Overall composition of $\textrm{C}_1$ ]{
\includegraphics[scale=0.6]{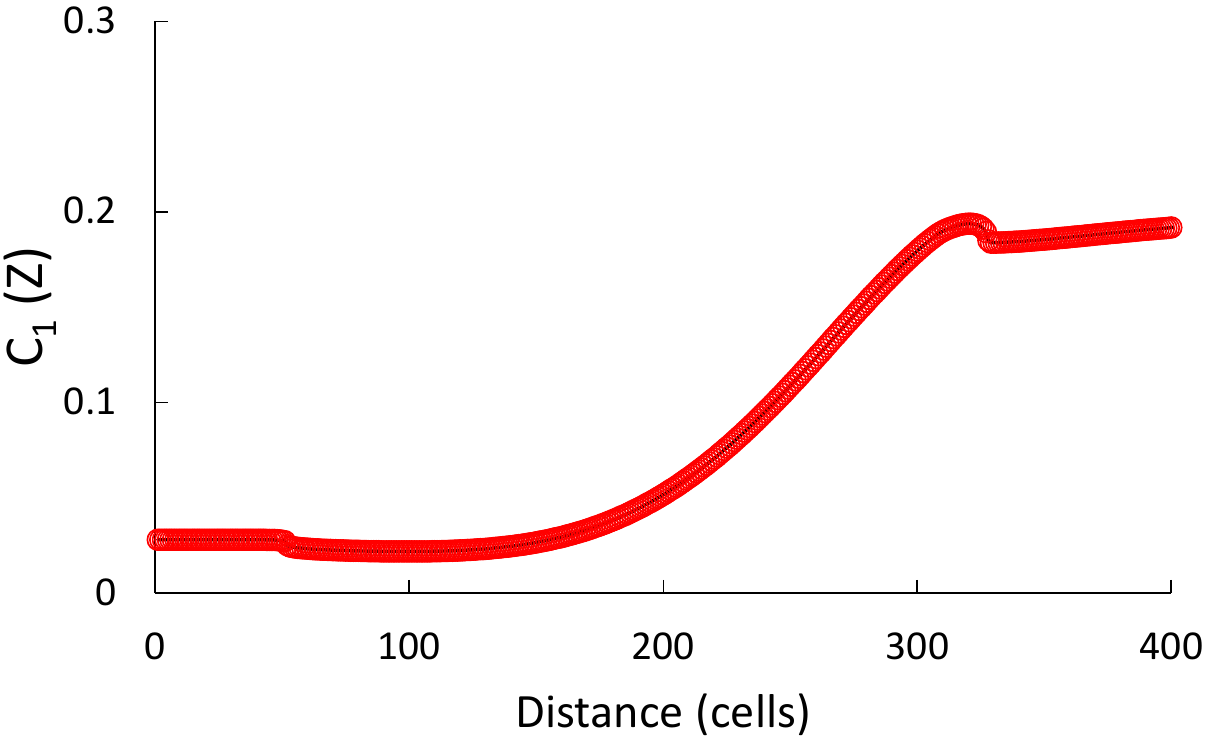}}
\\
\subfloat[Overall composition of $\textrm{CO}_2$ ]{
\includegraphics[scale=0.6]{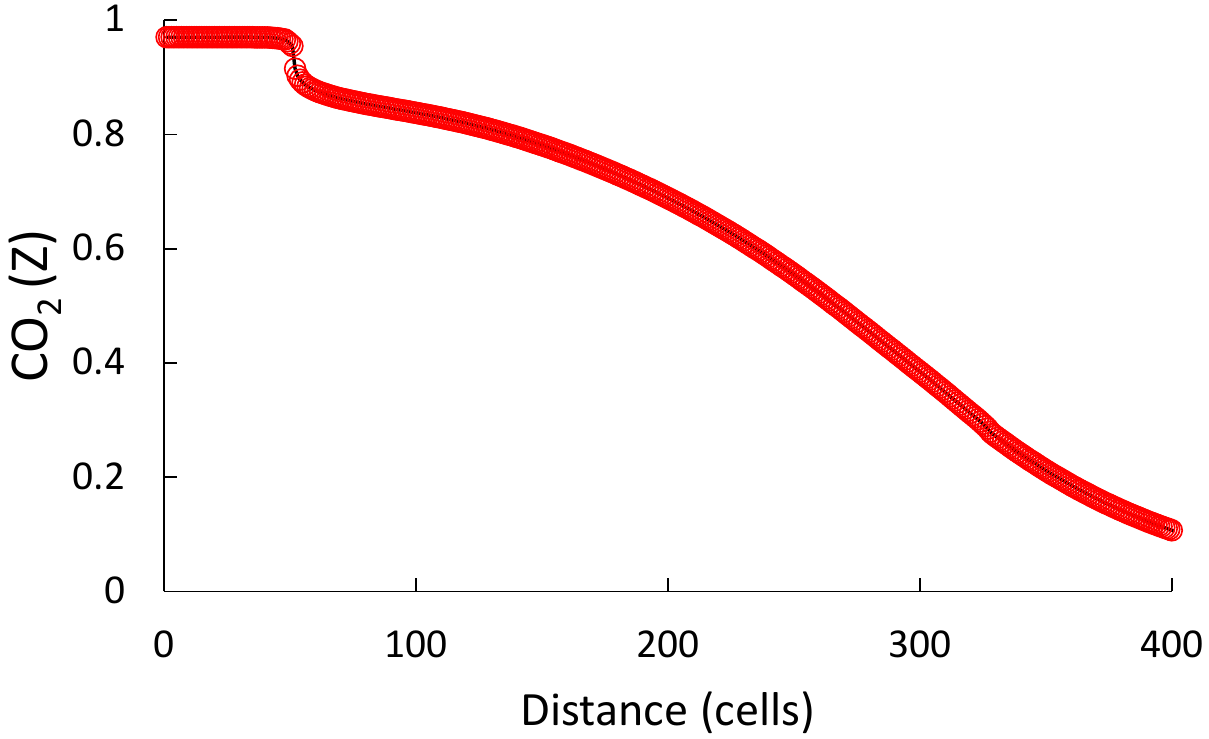}}
\caption{Gas saturation and overall composition profiles for the four-component fluid.}
\label{fig:4comp_1}
\end{figure}

The gas saturation and overall composition profiles are plotted in \textbf{Fig. \ref{fig:4comp_1}}. Note that the two formulations present the same solution profiles. 

\subsubsection{Case 2}

We test a gas-condensate system with initial compositions as $\left \{ 80 \%, 1 \%, 14 \%, 5 \% \right \}$ at an initial pressure of 100 bars and at a temperature 344 $\textrm{K}$. Production pressure is 95 bars. Injection pressure is 140 bars, and injection gas mixture is $\left \{ 1 \%, 97 \%, 1 \%, 1 \% \right \}$. The total simulation time is 500 days. 

%==================================================================

The nonlinear performance of Case 2 is summarized in Table \ref{tab:case_1d_4c_2}. The maxCFL is 60. A significant reduction in the iteration number is achieved by the smooth formulation.

\begin{table}[!htb]
\centering
\caption{Computational performance of Case 2 with the four-component fluid.}
\label{tab:case_1d_4c_2}
\begin{tabular}{|c|c|c|}
\hline
                                                           & Total iterations (Wasted) & Timesteps (Wasted) \\ \hline
\begin{tabular}[c]{@{}c@{}}Standard\\ Natural\end{tabular} & 248 (40)                    & 18 (2)             \\ \hline
\begin{tabular}[c]{@{}c@{}}Smooth \end{tabular}            & 90 (0)                     & 15 (0)             \\ \hline
\end{tabular}
\end{table}

\section{Results: SPE 10 model}

\begin{figure}[!htb]
\centering
\includegraphics[scale=0.61]{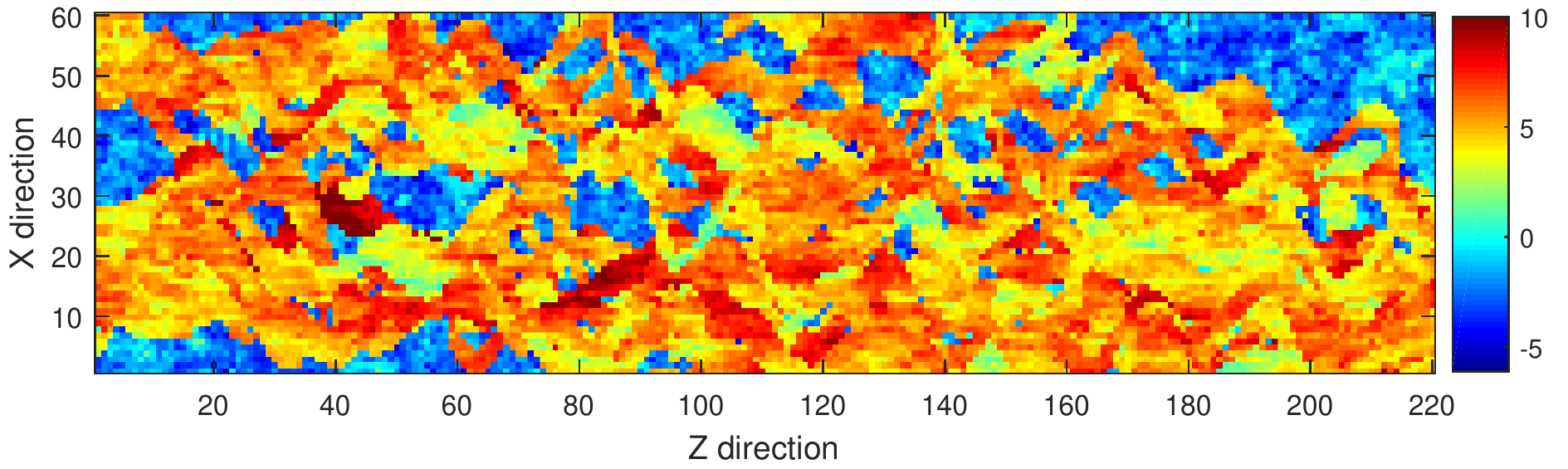}
\caption{Permeability (md) field of the SPE 10 model.}
\label{fig:perm_L}
\end{figure}

Permeability field of the bottom layer of the SPE 10 model is shown in \textbf{Fig. \ref{fig:perm_L}}. A uniform cell size 10 m is specified, and the porosity is 0.1. Positions of the producer and injector are (60, 1) and (1, 220), respectively. The relative permeabilities are quadratic for both phases.

\subsection{Case 1}

The fluid and compositions are the same as summarized in Table \ref{tab:3comp_fluid_1}. Initial pressure is 70 bars, at a temperature of 320 $\textrm{K}$. Production and injection pressures are 65 and 100 bars, respectively. The total simulation time is 400 days.

%==================================================================

The nonlinear performance of Case 1 is summarized in Table \ref{tab:case_2d_3c_1}. The maxCFL of the case is 88. The heterogeneous model is very challenging because of a large variation in the CFL numbers across the domain. Also frequent phase changes make the fluid displacement process highly nonlinear. As can be seen, the standard formulation suffers from many timestep cuts and wasted iterations. By comparison, the SBC method shows a smooth behavior, resulting in much improved nonlinear convergence. 

%--------------------------------------------------------------------------------

\begin{table}[!htb]
\centering
\caption{Computational performance of Case 1 with the SPE 10 model.}
\label{tab:case_2d_3c_1}
\begin{tabular}{|c|c|c|}
\hline
                                                           & Total iterations (Wasted) & Timesteps (Wasted) \\ \hline
\begin{tabular}[c]{@{}c@{}}Standard\\ Natural\end{tabular} & 560 (300)                 & 44 (15)             \\ \hline
\begin{tabular}[c]{@{}c@{}}SBC\\ $\epsilon_{min} = \textrm{1.0e-4}$\end{tabular}            & 193 (0)                   & 20 (0)             \\ \hline
\begin{tabular}[c]{@{}c@{}}SBC\\ $\epsilon_{min} = \textrm{1.0e-2}$\end{tabular}            & 171 (0)                   & 20 (0)             \\ \hline
\end{tabular}
\end{table}

%==================================================================

\begin{figure}[!htb]
\centering
\subfloat[Standard natural]{
\includegraphics[scale=0.4]{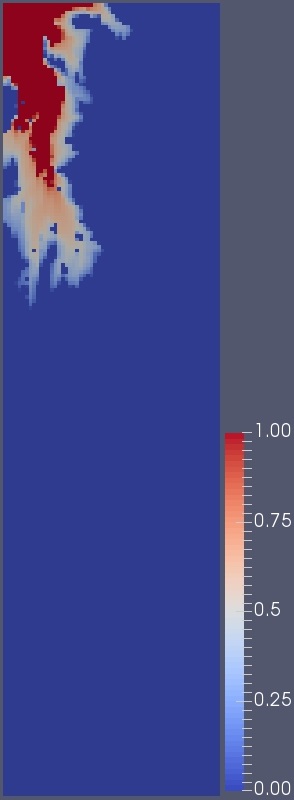}} \
\subfloat[SBC $\epsilon_{min} = \textrm{1.0e-4}$]{
\includegraphics[scale=0.4]{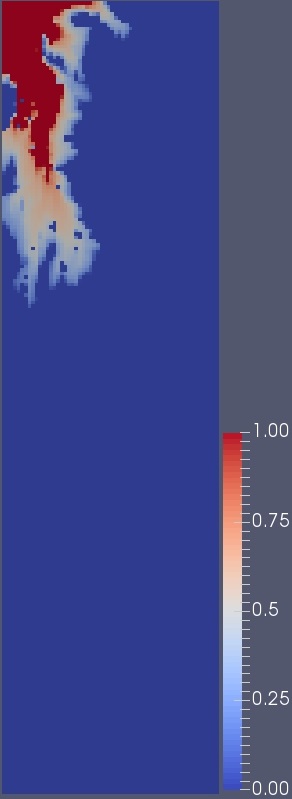}} \
\subfloat[SBC $\epsilon_{min} = \textrm{1.0e-2}$]{
\includegraphics[scale=0.4]{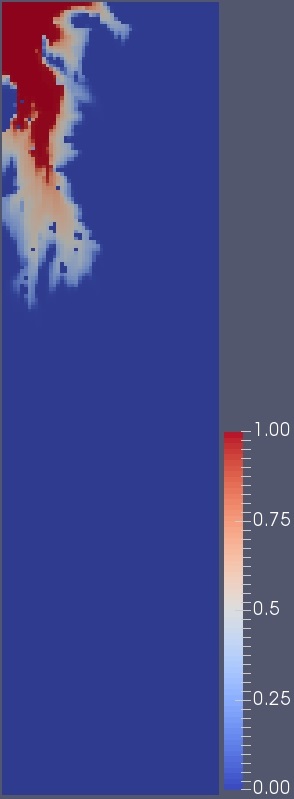}}
\caption{Gas saturation profiles for Case 1.}
\label{fig:SPE10_1_Sg}
\end{figure}

\begin{figure}[!htb]
\centering
\subfloat[Standard natural]{
\includegraphics[scale=0.4]{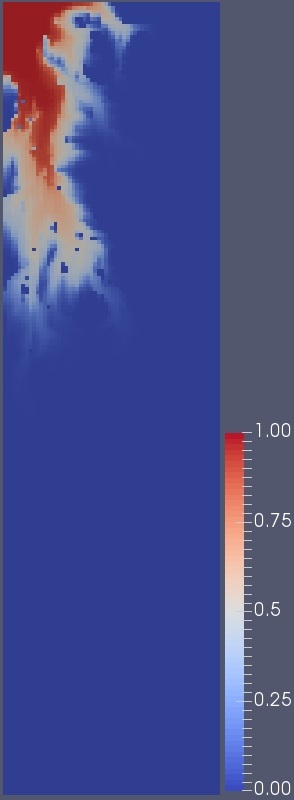}} \
\subfloat[SBC $\epsilon_{min} = \textrm{1.0e-4}$]{
\includegraphics[scale=0.4]{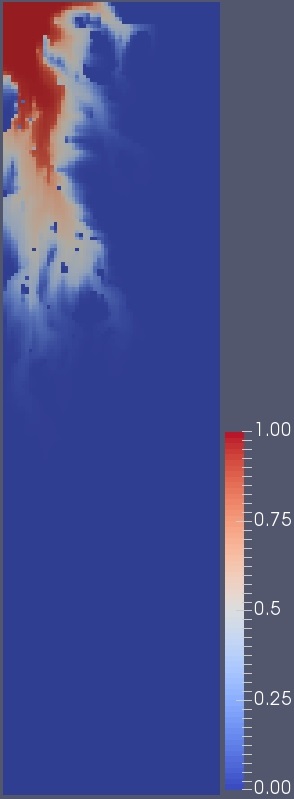}} \
\subfloat[SBC $\epsilon_{min} = \textrm{1.0e-2}$]{
\includegraphics[scale=0.4]{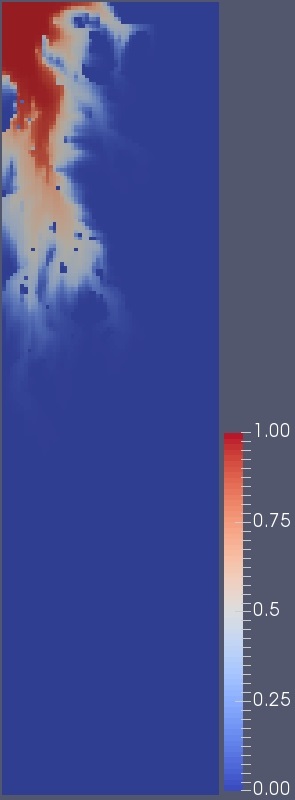}}
\caption{Overall composition profiles of $C_1$ for Case 1.}
\label{fig:SPE10_1_C1}
\end{figure}

The gas saturation and overall composition profiles are plotted in \textbf{Fig. \ref{fig:SPE10_1_Sg}} and \textbf{Fig. \ref{fig:SPE10_1_C1}}. The solutions from SBC closely matches the standard formulation, even for the level of $\epsilon_{min} = \textrm{1.0e-2}$.

\subsection{Case 2}

We also study a three-component system with $\left \{ K_1 = 2.5 , K_4 = 1.5, K_{10} = 0.3 \right \}$, and the other parameters remain unchanged.

The nonlinear performance of Case 2 is summarized in Table \ref{tab:case_2d_3c_2}. The maxCFL is 87. The iteration performance of the standard formulation becomes worse for the different $K$-values. Crossing phase boundaries along a single tie-line causes oscillations in the Newton solver. For multi-component systems, additional discontinuities can arise with the switches between key tie-lines. The SBC method provides smooth transitions across the entire phase boundaries, leading to superior global convergence performance.

\begin{table}[!htb]
\centering
\caption{Computational performance of Case 2 with the SPE 10 model.}
\label{tab:case_2d_3c_2}
\begin{tabular}{|c|c|c|}
\hline
                                                           & Total iterations (Wasted) & Timesteps (Wasted) \\ \hline
\begin{tabular}[c]{@{}c@{}}Standard\\ Natural\end{tabular} & 707 (360)                 & 50 (18)             \\ \hline
\begin{tabular}[c]{@{}c@{}}SBC\\ $\epsilon_{min} = \textrm{1.0e-4}$\end{tabular}            & 196 (0)                   & 20 (0)             \\ \hline
\begin{tabular}[c]{@{}c@{}}SBC\\ $\epsilon_{min} = \textrm{1.0e-2}$\end{tabular}            & 173 (0)                   & 20 (0)             \\ \hline
\end{tabular}
\end{table}

\subsection{Case 3}

We study the same fluid and $K$-values presented in Table \ref{tab:4comp_fluid_1}. Initial compositions are $\left \{ 80 \%, 0.1 \%, 14.9 \%, 5 \% \right \}$, at an initial pressure of 100 bars and at a temperature of 344 $\textrm{K}$. Production pressure is 95 bars. Injection pressure is 130 bars, and injection gas mixture is $\left \{ 0.98 \%, 99 \%, 0.01 \%, 0.01 \% \right \}$.

%==================================================================

The nonlinear performance of Case 3 is summarized in Table \ref{tab:case_2d_4c_3}. The maxCFL is 520. The SBC method exhibits favorable nonlinear convergence for this challenging case.

\begin{table}[!htb]
\centering
\caption{Computational performance of Case 3 with the SPE 10 model.}
\label{tab:case_2d_4c_3}
\begin{tabular}{|c|c|c|}
\hline
                                                           & Total iterations (Wasted) & Timesteps (Wasted) \\ \hline
\begin{tabular}[c]{@{}c@{}}Standard\\ Natural\end{tabular} & 356 (140)                 & 24 (7)             \\ \hline
\begin{tabular}[c]{@{}c@{}}SBC\\ $\epsilon_{min} = \textrm{1.0e-4}$\end{tabular}            & 110 (0)                   & 13 (0)             \\ \hline
\begin{tabular}[c]{@{}c@{}}SBC\\ $\epsilon_{min} = \textrm{1.0e-2}$\end{tabular}            & 96 (0)                   & 13 (0)             \\ \hline
\end{tabular}
\end{table}

%==================================================================

\begin{figure}[!htb]
\centering
\subfloat[Standard natural]{
\includegraphics[scale=0.4]{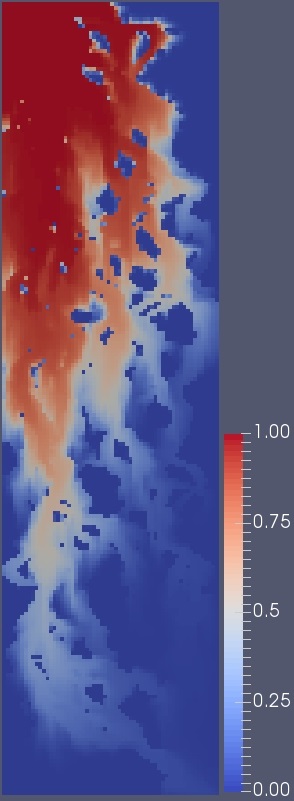}} \
\subfloat[SBC $\epsilon_{min} = \textrm{1.0e-4}$]{
\includegraphics[scale=0.4]{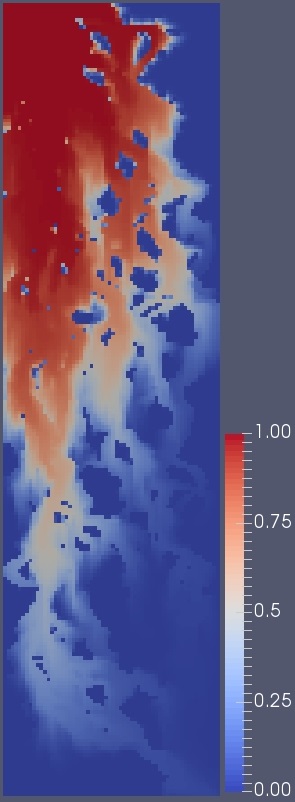}} \
\subfloat[SBC $\epsilon_{min} = \textrm{1.0e-2}$]{
\includegraphics[scale=0.4]{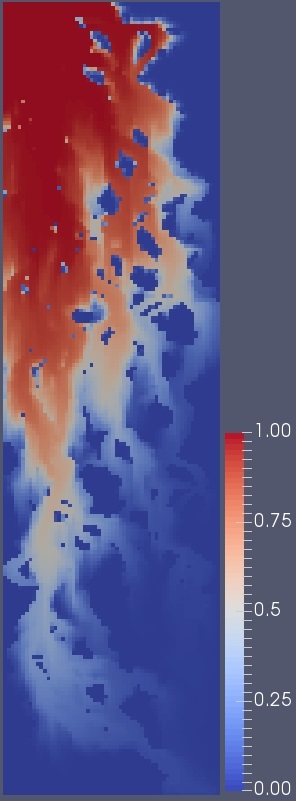}}
\caption{Overall composition profiles of $CO_2$ for Case 3.}
\label{fig:SPE10_3_CO2}
\end{figure}

The overall composition profiles are plotted in \textbf{Fig. \ref{fig:SPE10_3_CO2}}. Here the timestep size is reduced to 10 days, ensuring that the solutions from the different methods are compared with the same time-stepping schedule.

\section{Summary}

Frequent phase changes and oscillations around phase boundaries can cause severe convergence problems during compositional simulations. The objective of this work is to develop a smooth formulation that removes all the property switches and discontinuities associated with phase changes. Here we first reformulate the coupled system, so that the discontinuities are transferred to the phase equilibrium model. A single and concise non-smooth equation is achieved and then a smoothing approximation can be made. The reformulation is based on a mixed complementarity problem (MCP) proposed for the phase equilibrium. The developed formulation with a smoothing parameter provides smooth transitions of variables across all the phase regimes. We also employ a continuation method with the smooth system as a homotopy mapping. 

We evaluate the efficiency of the new formulation and the continuation method using several complex problems. For most of the cases, the standard natural formulation suffers from multiple timestep cuts and subsequent wasted Newton iterations. In contrast, the developed formulation and method exhibit superior global convergence, requiring no timestep cut. The SBC method shows a negligible impact on solution accuracy, while providing smooth and stable iterative performance. Moreover, SBC works robustly for a wide range of flow conditions without parameter tuning. 

In this work, we mainly focus on $K$-values to evaluate the phase equilibrium.~Preliminary tests are conducted for an EoS-based compositional problem.~The smooth formulation can be readily applied to the black-oil fluid model (with dissolved gas and vaporized oil).~Our first application target is on real-field black-oil models.

\section*{Acknowledgements}

This work was supported by the Chevron/Schlumberger INTERSECT Research \& Prototyping project. The authors thank Chevron for permission to publish the paper. Petroleum Research Institute for Reservoir Simulation (SUPRI-B) at Stanford University is gratefully acknowledged for providing the AD-GPRS platform.

\section*{Appendix A. Formulation with complementarity conditions}

%  The semismooth method is basically the same as the standard Newton method in all regions where the functions are differentiable.  at the other points, the values of the derivatives are defined as one of the directional derivatives. 

Lauser et al. (2011) proposed a formulation based on complementarity conditions for handling phase changes. For each phase, the sum of the molar fractions is bounded from above by one, with equality holding if the phase is present, 
\begin{equation} 
\sum_{c=1}^{n_c} x_{c,l} \leqslant 1 \ , \quad \sum_{c=1}^{n_c} x_{c,l} = 1 \ \ \textrm{if phase} \ l \ \textrm{is present.}
\end{equation}
The corresponding complementarity conditions are given as, 
\begin{equation}
1 - \sum_{c=1}^{n_c} x_{c,l} \geqslant 0 \ , \ \ \nu_l \geqslant 0 \ , \ \ \nu_l \left ( 1 - \sum_{c=1}^{n_c} x_{c,l} \right ) = 0 
\end{equation}

The complementarity conditions with the inequalities, can be reformulated equivalently as non-smooth equations, so that the phase equilibrium system becomes, 
\begin{equation} 
y_c - K_c x_c = 0 ,
\end{equation}

\begin{equation} 
z_c - \nu_o x_{c} - \left( 1 - \nu_o \right) y_{c} = 0 ,
\end{equation}

\begin{equation} 
\textrm{min} \left \{ \nu_l , \left ( 1 - \sum_{c=1}^{n_c} x_{c,l} \right ) \right \} = 0.
\end{equation}
The minimum function represents the nonlinear complementarity function.

The above formulation may produce the solutions with negative molar fractions, especially if smoothing approximations are applied. This limitation can cause severe numerical issues for compositional problems of practical interest.

In Lauser et al. (2011), a semismooth Newton method with locally superlinear convergence is employed to solve the non-smooth system. However, the semismooth method may exhibit poor global convergence: the iterations may diverge, when the starting point is not close enough to a solution.

\section*{Appendix B. Discretization methods}

A standard finite-volume scheme is applied as the spatial discretization for the mass conservation equations. A two-point flux approximation (TFPA) is used to approximate the flux across a cell interface. The method of choice for the time discretization is the fully-implicit scheme. The discrete form of conservation equation is given as, 
\begin{equation} 
\frac{V}{\Delta t} \left [ \left ( \phi \rho_T z_c \right )^{n+1} - \left ( \phi \rho_T z_c \right )^{n} \right ] - \sum_{ij}\left ( x_{c} \rho_o F_o + y_{c} \rho_g F_g \right )^{n+1} - Q_c^{n+1} = 0.
\end{equation}
where superscripts denote timesteps, and $\Delta t$ is the timestep size. $V$ is the cell volume. All indexes related to the cell numeration are neglected. The accumulation term involves the total density, 
\begin{equation} 
\rho_T z_c = x_{c} \rho_o s_o + y_{c} \rho_g s_g 
\end{equation}
and
\begin{equation} 
\rho_T = \begin{cases}
s_o \rho_o (\textbf{x}) + s_g \rho_g (\textbf{y}) , \ \ \textrm{two phase}, \\ 
\rho_l (\textbf{z}) , \ \ \textrm{one phase}.
\end{cases}
\end{equation}
where $\rho_o (\textbf{x})$ indicates the density computed at a composition $\textbf{x}$, and $\rho_l (\textbf{z})$ is the density computed in the single-phase regime at a composition $\textbf{z}$.

The discrete phase flux across the interface $(ij)$ between two cells is written as, 
\begin{equation} 
F_{l,ij} = \Upsilon_{ij} \lambda_{l,ij} \Delta \Phi_{l,ij}
\end{equation}
where subscript $(ij)$ denotes quantities defined at the cell interface. $\Upsilon_{ij}$ is the interface transmissibility. $\Delta \Phi_{l,ij} = \Delta p_{ij} - g_{l,ij}$ is the phase potential difference with the discrete weights $g_{l,ij} = \rho_{l,ij} \ g \Delta h_{ij}$. The phase and compositional coefficients associated with the flux terms are evaluated using the Phase-Potential Upwinding (PPU) scheme.

\section*{Appendix C. Simulation results on EoS-based problems}

We provide preliminary test results that demonstrate the applicability of the smooth compositional formulation to EoS-based problems.~The Peng-Robinson EoS model is used. The reservoir models include: (a) homogeneous 1D model; (b) SPE 10 model.

We note that although the new formulation can be directly applied, additional challenges could arise within super-critical regions of compositional space. In practice, improvements for the stability analysis and flash procedures may be necessary to ensure robustness. Modeling of complex miscible displacements is subject to future research.

\subsection*{Results: 1D model}

% unless otherwise stated
% The simulation example presented previously is still employed in this section. 

\begin{figure}[!htb]
\centering
\subfloat[Gas saturation]{
\includegraphics[width=8cm,height=4cm]{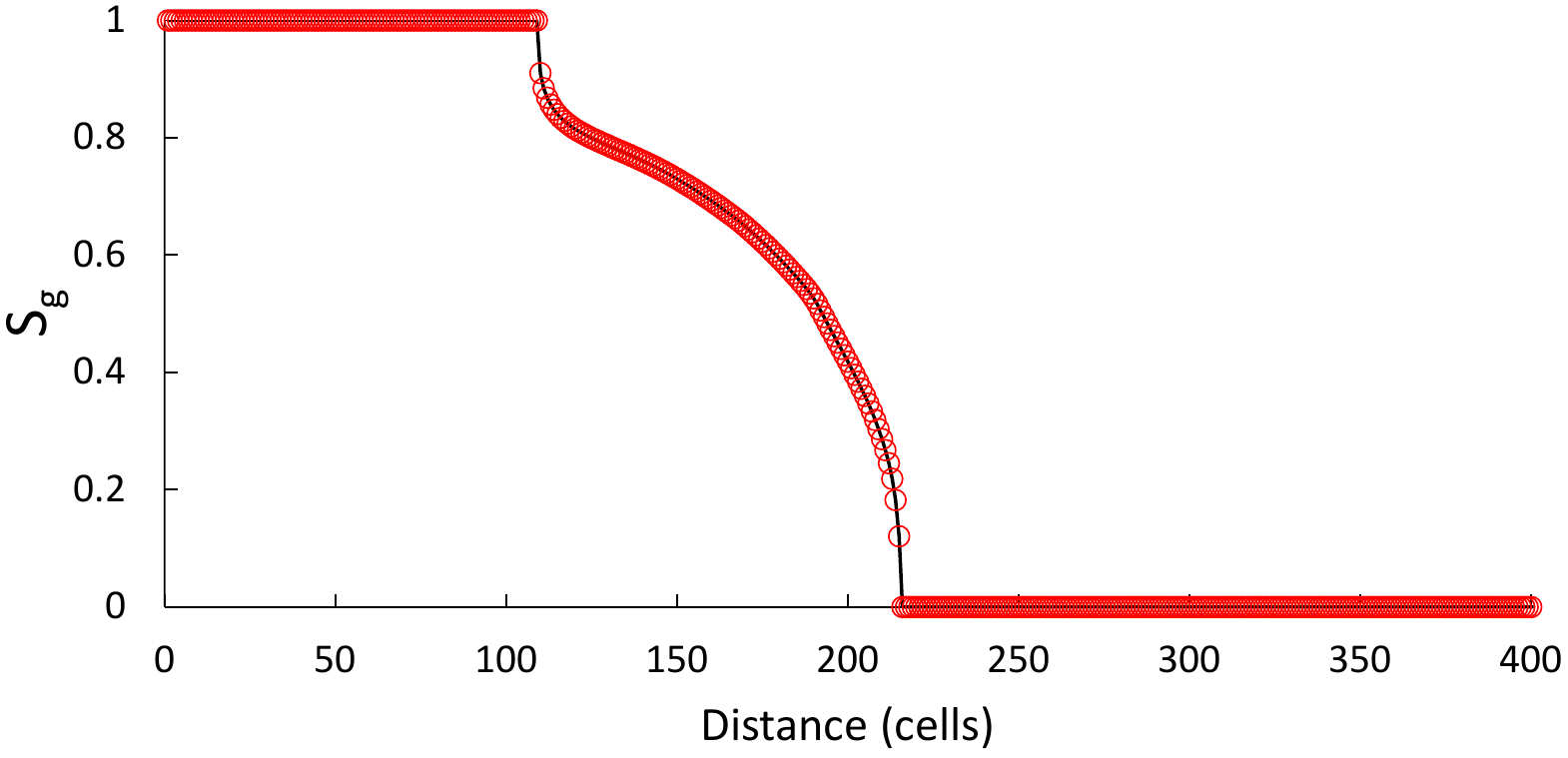}}
\\
\subfloat[Overall composition of $\textrm{C}_1$ ]{
\includegraphics[width=8cm,height=4cm]{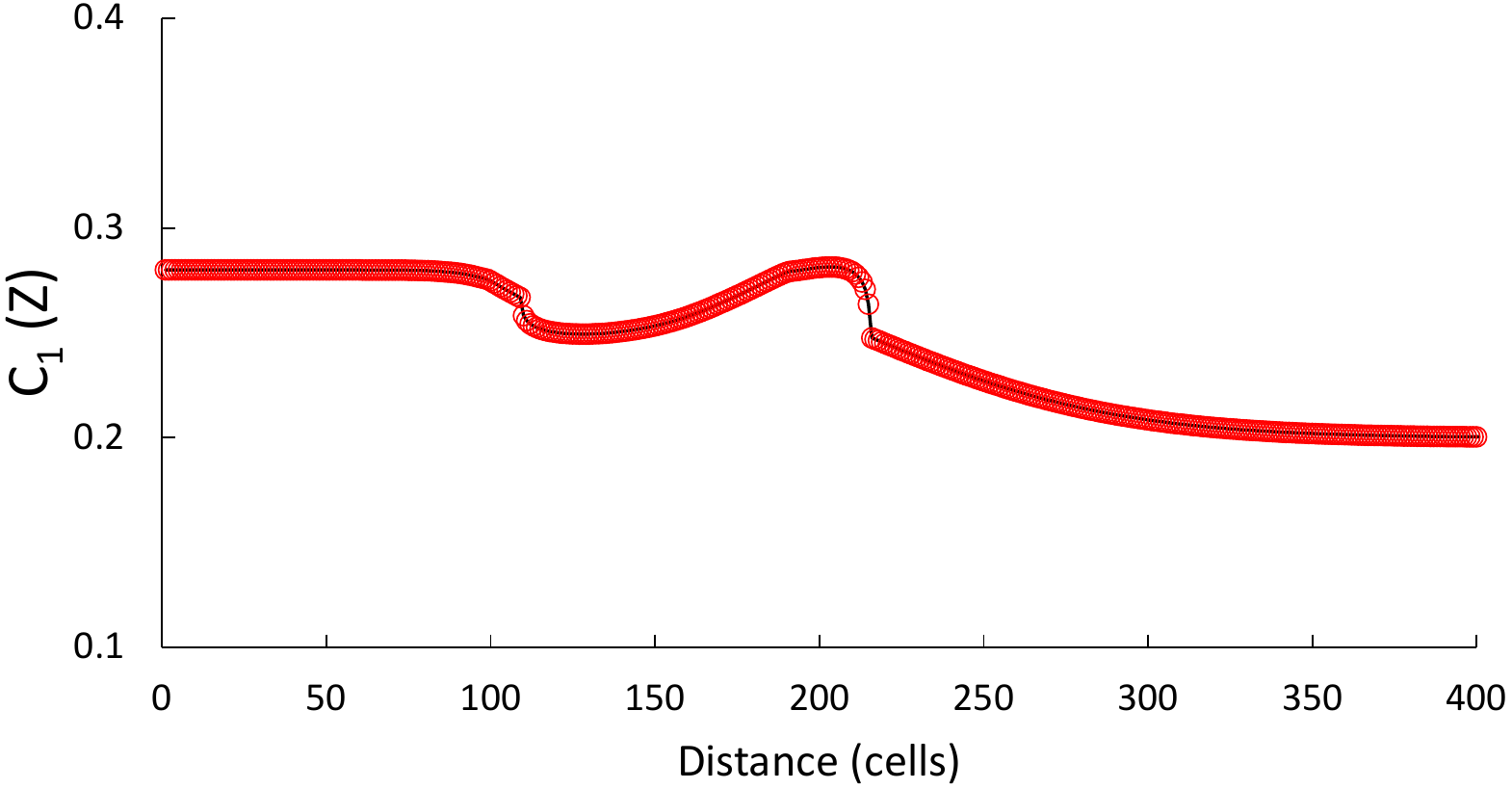}}
\\
\subfloat[Overall composition of $\textrm{CO}_2$ ]{
\includegraphics[width=8cm,height=4cm]{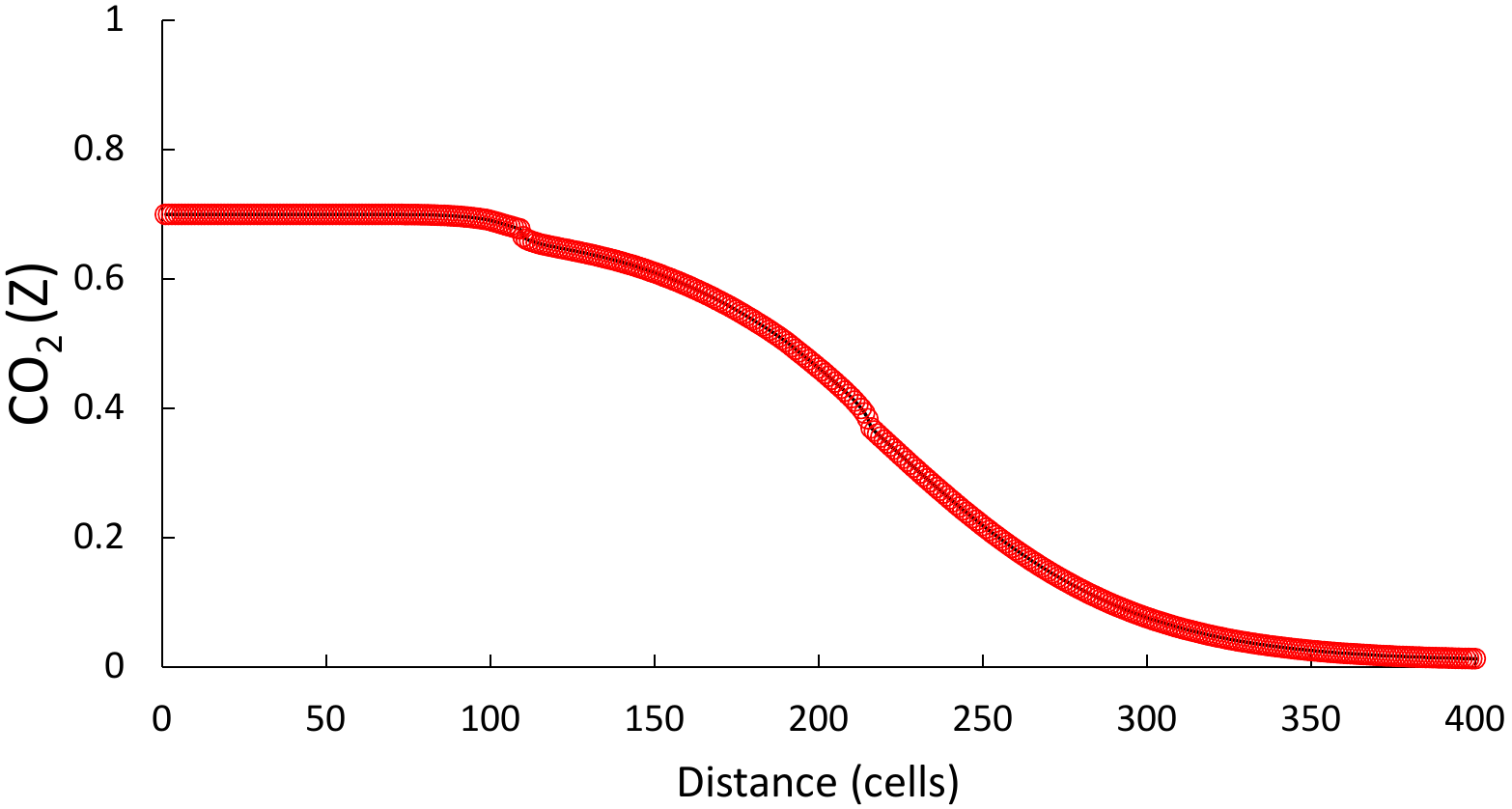}}
\caption{Gas saturation and overall composition profiles of the EoS-based 1D model.}
\label{fig:eos_4comp_1}
\end{figure}

We first validate the efficiency of the new nonlinear solver on the 1D model specified in Table \ref{tab:specification_m}. The four-component fluid system $\left \{ \textrm{C}_1 , \textrm{CO}_2 , \textrm{C}_4 , \textrm{C}_{10} \right \}$ is used.~The initial compositions are $\left \{ 20 \%, 1 \%, 29 \%, 50 \% \right \}$, at an initial pressure of 80 bars and at a temperature 373 $\textrm{K}$.~The injection pressure is 190 bars, and the injection gas mixture is $\left \{ 28 \%, 70 \%, 1 \%, 1 \% \right \}$.~The gas saturation and overall composition profiles are plotted in \textbf{Fig.~\ref{fig:eos_4comp_1}}. Comparison of the nonlinear results between the standard and smoothing-based continuation methods is summarized in Table \ref{tab:case_eos_1}.

% MaxCFL averaged by timesteps = 1.958e+01

\begin{table}[!htb]
\centering
\caption{Computational performance of the EoS-based 1D model.}
\label{tab:case_eos_1}
\begin{tabular}{|c|c|c|}
\hline
                                                           & Total iterations (Wasted) & Timesteps (Wasted) \\ \hline
\begin{tabular}[c]{@{}c@{}}Standard\\ Natural\end{tabular} & 842 (435)                 & 29 (22)             \\ \hline
\begin{tabular}[c]{@{}c@{}}SBC\\ $\epsilon_{min} = \textrm{1.0e-4}$\end{tabular}            & 192 (0)                   & 16 (0)             \\ \hline
\end{tabular}
\end{table}

\subsection*{Results: SPE 10 model}

We run a test on the SPE 10 model with the four-component fluid.~The initial compositions are $\left \{ \textrm{C}_1 (80 \%), \textrm{CO}_2 (0 \%), \textrm{C}_4 (15 \%), \textrm{C}_{10} (5 \%) \right \}$, at an initial pressure of 101 bars and at a temperature 344 $\textrm{K}$. The injection pressure is 130 bars, and the injection fluid is a two-component mixture $\left \{ \textrm{C}_1 (1 \%), \textrm{CO}_2 (99 \%) \right \}$.

\begin{figure}[!htb]
\centering
\subfloat[Gas saturation]{
\includegraphics[scale=0.4]{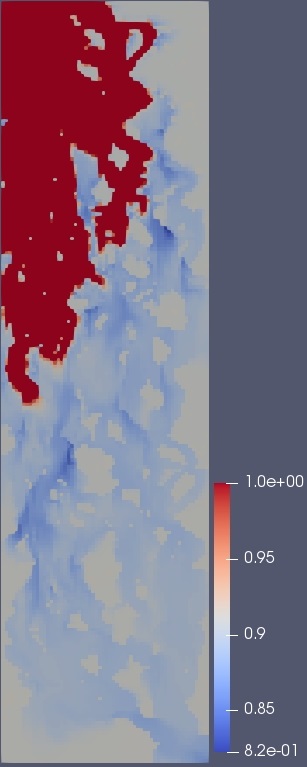}}
\subfloat[$\textrm{CO}_2$ ]{
\ \
\includegraphics[scale=0.4]{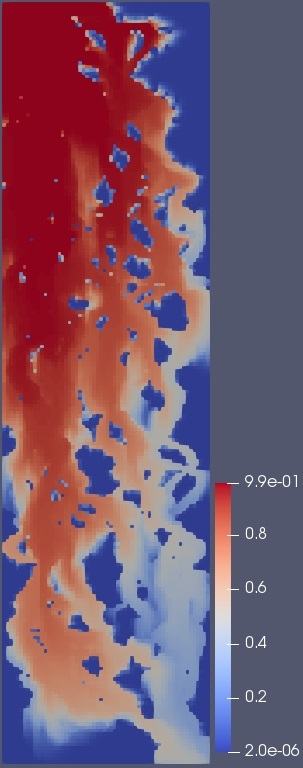}}
\caption{Gas saturation and overall composition profiles of the EoS-based SPE 10 model.}
\label{fig:SPE10_Jessen}
\end{figure}

The gas saturation and overall composition profiles are plotted in \textbf{Fig.~\ref{fig:SPE10_Jessen}}.~The nonlinear results are summarized in Table \ref{tab:case_eos_2}.~We can see that the standard formulation suffers from a large number of timestep cuts and wasted iterations. In contrast, the SBC method significantly improves the nonlinear convergence performance.

%-----------------------------------------------------------------------------------

% MaxCFL averaged by timesteps = 4.998e+02

\begin{table}[!htb]
\centering
\caption{Computational performance of the EoS-based SPE 10 model.}
\label{tab:case_eos_2}
\begin{tabular}{|c|c|c|}
\hline
                                                           & Total iterations (Wasted) & Timesteps (Wasted) \\ \hline
\begin{tabular}[c]{@{}c@{}}Standard\\ Natural\end{tabular} & 386 (180)                 & 17 (9)             \\ \hline
\begin{tabular}[c]{@{}c@{}}SBC\\ $\epsilon_{min} = \textrm{1.0e-4}$\end{tabular}            & 138 (0)                   & 12 (0)             \\ \hline
\end{tabular}
\end{table}

\section*{References}

Aziz, K., Settari, A., 1979. Petroleum Reservoir Simulation. Chapman \& Hall.

Acs, G., Doleschall, S. and Farkas, E., 1985. General purpose compositional model. Society of Petroleum Engineers Journal, 25(04), pp.543-553.

Abadpour, A. and Panfilov, M., 2009. Method of negative saturations for modeling two-phase compositional flow with oversaturated zones. Transport in porous media, 79(2), pp.197-214.

Alpak, F.O., 2010. A mimetic finite volume discretization method for reservoir simulation. SPE Journal, 15(02), pp.436-453.

Alpak, F.O. and Vink, J.C., 2018. A Variable-Switching Method for Mass-Variable-Based Reservoir Simulators. SPE Journal, 23(05), pp.1-469.

Bullard, L.G. and Biegler, L.T., 1993. Iterated linear programming strategies for non-smooth simulation: A penalty based method for vapor—liquid equilibrium applications. Computers $\&$ chemical engineering, 17(1), pp.95-109.

Coats, K.H., 1980. An equation of state compositional model. Society of Petroleum Engineers Journal, 20(05), pp.363-376.

Collins, D.A., Nghiem, L.X., Li, Y.K. and Grabonstotter, J.E., 1992. An efficient approach to adaptive-implicit compositional simulation with an equation of state. SPE reservoir engineering, 7(02), pp.259-264.

Chen, B. and Harker, P.T., 1993. A non-interior-point continuation method for linear complementarity problems. SIAM Journal on Matrix Analysis and Applications, 14(4), pp.1168-1190.

Cao, H., 2002. Development of techniques for general purpose simulators (Doctoral dissertation, Stanford University).

Deuflhard, P., 2011. Newton methods for nonlinear problems: affine invariance and adaptive algorithms (Vol. 35). Springer Science $\&$ Business Media.

ECLIPSE Technical Description 2008. Houston: Schlumberger GeoQuest. 

Gopal, V. and Biegler, L.T., 1997. Nonsmooth dynamic simulation with linear programming based methods. Computers \& chemical engineering, 21(7), pp.675-689.

Gharbia, I.B., Flauraud, E. and Michel, A., 2015, February. Study of compositional multi-phase flow formulations with cubic eos. In SPE Reservoir Simulation Symposium. Society of Petroleum Engineers.

Hamon, F.P., Mallison, B.T. and Tchelepi, H.A., 2016. Implicit Hybrid Upwind scheme for coupled multiphase flow and transport with buoyancy. Computer Methods in Applied Mechanics and Engineering, 311, pp.599-624.

Jenny, P., Tchelepi, H.A. and Lee, S.H., 2009. Unconditionally convergent nonlinear solver for hyperbolic conservation laws with S-shaped flux functions. Journal of Computational Physics, 228(20), pp.7497-7512.

Jiang, J. and Younis, R.M., 2017. Efficient C1-continuous phase-potential upwind (C1-PPU) schemes for coupled multiphase flow and transport with gravity. Advances in Water Resources, 108, pp.184-204.

Jiang, J. and Tchelepi, H.A., 2018. Dissipation-based continuation method for multiphase flow in heterogeneous porous media. Journal of Computational Physics, 375, pp.307-336.

Kanzow, C., 1996. Some noninterior continuation methods for linear complementarity problems. SIAM Journal on Matrix Analysis and Applications, 17(4), pp.851-868.

Knoll, D.A. and Keyes, D.E., 2004. Jacobian-free Newton–Krylov methods: a survey of approaches and applications. Journal of Computational Physics, 193(2), pp.357-397.

Lohrenz, J., Bray, B.G. and Clark, C.R., 1964. Calculating viscosities of reservoir fluids from their compositions. Journal of Petroleum Technology, 16(10), pp.1-171.

Li, D. and Fukushima, M., 2000. Smoothing Newton and quasi-Newton methods for mixed complementarity problems. Computational Optimization and Applications, 17(2-3), pp.203-230.

Lauser, A., Hager, C., Helmig, R. and Wohlmuth, B., 2011. A new approach for phase transitions in miscible multi-phase flow in porous media. Advances in Water Resources, 34(8), pp.957-966.

Lee, S.H., Efendiev, Y. and Tchelepi, H.A., 2015. Hybrid upwind discretization of nonlinear two-phase flow with gravity. Advances in Water Resources, 82, pp.27-38.

Lee, S.H. and Efendiev, Y., 2016. C1-Continuous relative permeability and hybrid upwind discretization of three phase flow in porous media. Advances in Water Resources, 96, pp.209-224.

Michelsen, M.L., 1982a. The isothermal flash problem. Part I. Stability. Fluid phase equilibria, 9(1), pp.1-19.

Michelsen, M.L., 1982b. The isothermal flash problem. Part II. Phase-split calculation. Fluid phase equilibria, 9(1), pp.21-40.

Orr, F.M., 2007. Theory of gas injection processes (Vol. 5). Copenhagen: Tie-Line Publications.

Sahlodin, A.M., Watson, H.A. and Barton, P.I., 2016. Nonsmooth model for dynamic simulation of phase changes. AIChE Journal, 62(9), pp.3334-3351.

Voskov, D.V. and Tchelepi, H.A., 2009. Compositional space parameterization: theory and application for immiscible displacements. SPE Journal, 14(03), pp.431-440.

Voskov, D.V. and Tchelepi, H.A., 2011, January. Compositional nonlinear solver based on trust regions of the flux function along key tie-lines. In SPE Reservoir Simulation Symposium. Society of Petroleum Engineers.

Voskov, D.V. and Tchelepi, H.A., 2012. Comparison of nonlinear formulations for two-phase multi-component EoS based simulation. Journal of Petroleum Science and Engineering, 82, pp.101-111.

Voskov, D., 2012. An extended natural variable formulation for compositional simulation based on tie-line parameterization. Transport in porous media, 92(3), pp.541-557.

Whitson, C.H. and Michelsen, M.L., 1989. The negative flash. Fluid Phase Equilibria, 53, pp.51-71.

Wong, T.W., Firoozabadi, A. and Aziz, K., 1990. Relationship of the volume-balance method of compositional simulation to the Newton-Raphson method. SPE Reservoir Engineering, 5(03), pp.415-422.

Wang, X. and Tchelepi, H.A., 2013. Trust-region based solver for nonlinear transport in heterogeneous porous media. Journal of Computational Physics, 253, pp.114-137.

Younis, R. and Aziz, K., 2007, January. Parallel automatically differentiable data-types for next-generation simulator development. In SPE Reservoir Simulation Symposium. Society of Petroleum Engineers.

Younis, R., Tchelepi, H.A. and Aziz, K., 2010. Adaptively Localized Continuation-Newton Method--Nonlinear Solvers That Converge All the Time. SPE Journal, 15(02), pp.526-544.

Yan, W., Belkadi, A., Stenby, E.H. and Michelsen, M., 2013. Study on the application of the tie-line-table-look-up-based methods to flash calculations in compositional simulations. SPE Journal, 18(05), pp.932-942.

Zhou, Y., Tchelepi, H.A. and Mallison, B.T., 2011, January. Automatic differentiation framework for compositional simulation on unstructured grids with multi-point discretization schemes. In SPE Reservoir Simulation Symposium. Society of Petroleum Engineers.

Zaydullin, R., Voskov, D. and Tchelepi, H.A., 2012. Nonlinear formulation based on an equation-of-state free method for compositional flow simulation. SPE Journal, 18(02), pp.264-273.

\end{document}